\documentclass[aps, prd, onecolumn, tightenlines, notitlepage, superscriptaddress, nofootinbib, preprintnumbers, floatfix,showkeys,11pt]{revtex4-2}

\usepackage[normalem]{ulem}
\usepackage{amstext}
\usepackage{amssymb}
\usepackage{amsmath}
\usepackage{graphicx}
\graphicspath{{plots/}}
\usepackage{url}
\usepackage{color}
\usepackage{ulem}
\usepackage[utf8]{inputenc}
\pdfoutput=1
\usepackage{textcomp}
\usepackage{makecell}\usepackage{booktabs,siunitx,array,threeparttable}
\usepackage{epsfig,amsfonts,mathrsfs,amsmath,amssymb,graphicx,color,slashed,multirow}

\usepackage[x11names]{xcolor}
\usepackage[colorlinks]{hyperref}

\usepackage{textgreek} % To add greek letters to the text
\usepackage{caption, subcaption} % Better figure customization
\usepackage{booktabs} % Cool tables

\usepackage{xspace}
\def\cevns{CE$\nu$NS\xspace}
\def\eves{E$\nu$ES\xspace}

\definecolor{vdrgreen}{rgb}{0.0, 0.7, 0.0}

\definecolor{lightapricot}{rgb}{0.99, 0.84, 0.69}
\definecolor{nicered}{rgb}{0.7,0.1,0.1}
\definecolor{nicegreen}{rgb}{0.1,0.5,0.1}
\definecolor{coral}{rgb}{1.0, 0.5, 0.31}
\definecolor{blue(ncs)}{rgb}{0.0, 0.53, 0.74}
\definecolor{darkspringgreen}{rgb}{0.09, 0.45, 0.27}
\definecolor{seagreen}{rgb}{0.18, 0.55, 0.34}
\definecolor{cadmiumgreen}{rgb}{0.0, 0.42, 0.24}
\definecolor{chromeyellow}{rgb}{1.0, 0.65, 0.0}
\definecolor{darkturquoise}{rgb}{0.0, 0.81, 0.82}
\definecolor{denim}{rgb}{0.08, 0.38, 0.74}
\definecolor{purple(x11)}{rgb}{0.63, 0.36, 0.94}
\definecolor{red(ncs)}{rgb}{0.77, 0.01, 0.2}
\definecolor{ruddypink}{rgb}{0.88, 0.56, 0.59}
\definecolor{slateblue}{rgb}{0.42, 0.35, 0.8}
\definecolor{airforceblue}{rgb}{0.36, 0.54, 0.66}
\definecolor{orange(colorwheel)}{rgb}{1.0, 0.5, 0.0}

\makeatletter
    \newcommand{\colorboxed}[3][white]{\fcolorbox{#2}{#1}{\m@th$\displaystyle#3$}}
\makeatother

\AtBeginDocument{\hypersetup{citecolor=seagreen,linkcolor=seagreen,urlcolor=seagreen}}
\usepackage{appendix}

\begin{document}

\title{{\LARGE Testing neutrino electromagnetic properties at current and future dark matter experiments}}

\author{Carlo Giunti}
\email{carlo.giunti@to.infn.it}
\affiliation{Istituto Nazionale di Fisica Nucleare (INFN), Sezione di Torino, Via P. Giuria 1, I--10125 Torino, Italy}

\author{Christoph A. Ternes}
\email{christoph.ternes@lngs.infn.it}
%\affiliation{Istituto Nazionale di Fisica Nucleare (INFN), Sezione di Torino, Via P. Giuria 1, I--10125 Torino, Italy}
%\affiliation{Dipartimento di Fisica, Universit\`a di Torino, via P. Giuria 1, I--10125 Torino, Italy}
\affiliation{Istituto Nazionale di Fisica Nucleare (INFN), Laboratori Nazionali del Gran Sasso, 67100 Assergi, L’Aquila (AQ), Italy
}

\begin{abstract}
We analyze data from the dark matter direct detection experiments PandaX-4T, LUX-ZEPLIN and XENONnT to place bounds on  neutrino electromagnetic properties
(magnetic moments, millicharges, and charge radii). We also show how these bounds will improve at the future facility DARWIN. In our analyses we implement a more conservative treatment of background uncertainties than usually done in the literature. From the combined analysis of all three experiments we can place very strong bounds on the neutrino magnetic moments and on the neutrino millicharges. We show that even though the bounds on the neutrino charge radii are not very strong from the analysis of current data, DARWIN could provide the first measurement of the electron neutrino charge radius, in agreement with the Standard Model prediction.
\end{abstract}
\maketitle

\section{Introduction}
The investigation of neutrino properties is one of the most active research fields in particle physics. In the Standard Model neutrinos are massless particles which interact only via weak interactions. Through the observation of neutrino oscillations we know, however, that at least two neutrinos are massive particles. Therefore, the Standard Model has to be extended to account for neutrino masses.  

In some of the extensions of the Standard Model neutrinos can acquire electromagnetic properties through quantum loop effects. Therefore, some models of physics beyond the Standard Model predict the interaction of neutrinos with electromagnetic fields and electromagnetic interactions with charged particles.
Moreover, even the Standard Model predicts non-zero neutrino charge radii due to radiative corrections.
For detailed reviews on the theories of neutrino electromagnetic properties we refer the reader to Refs.~\cite{Giunti:2014ixa,Giunti:2015gga}.

In this paper we test the neutrino electromagnetic properties
(magnetic moments, millicharges, and charge radii)
by analyzing data from dark matter direct detection experiments. These experiments aim to measure the nuclear or electron recoils from dark matter interacting with the material in the detector, which is Xenon for all experiments under consideration in this work. In this paper we consider the data from the PandaX-4T experiment~\cite{PandaX:2022ood}, from LUX-ZEPLIN (LZ)~\cite{LZ:2022lsv}, and from XENONnT~\cite{XENON:2022ltv}. We also show the sensitivity of the future experiment DARWIN~\cite{DARWIN:2020bnc}.

One of the background sources for the dark matter search in these experiments is the elastic scattering of solar neutrinos on the electrons in the detector material. Therefore, if some new physics model changes the standard electron-neutrino elastic scattering (E$\nu$ES) cross section, it can be tested at dark matter direct detection experiments. One possibility to alter this process is the presence of neutrino electromagnetic properties. 

Our paper is structured as follows: In Section~\ref{sec:theory} we discuss how neutrino magnetic moments, neutrino millicharges and neutrino charge radii alter the E$\nu$ES process. In Section~\ref{sec:analysis} we detail the analysis procedure of the experiments under consideration. We proceed to present and discuss our results in Section~\ref{sec:res}, before concluding in Section~\ref{sec:conclusions}.

\section{Theoretical framework}
\label{sec:theory}

In this paper we obtain bounds on neutrino electromagnetic properties
from the data of Xenon dark matter experiments through
the elastic scattering of solar neutrinos with electrons (\eves).
These experiments are sensitive to the neutrino electromagnetic properties
through their contributions to \eves in addition to the
Standard Model \eves cross section.
Therefore, we first present the Standard Model \eves cross section
in Subsection~\ref{sub:SMcs},
and then we discuss the cross sections due to the neutrino electromagnetic properties that we consider:
magnetic moments in Subsection~\ref{sub:MMcs},
electric charges in Subsection~\ref{sub:ECcs}, and
charge radii in Subsection~\ref{sub:CRcs}.
These three electromagnetic properties
are the observable effective electromagnetic properties
of ultrarelativistic neutrinos,
for which the effective magnetic moments include possible electric moments
and the effective charge radii include possible anapole moments~\cite{Giunti:2014ixa,Giunti:2015gga,Giunti:2022aea}.

Solar neutrinos oscillate and arrive at a detector on Earth
as a mixture of
$\nu_{e}$, $\nu_{\mu}$, and $\nu_{\tau}$,
whose fluxes are given by
\begin{equation}
\Phi_{\nu_{e}}^{i}
=
\Phi_{\nu_{e}}^{i\,\odot} P_{ee},
\quad
\Phi_{\nu_{\mu}}^{i}
=
\Phi_{\nu_{e}}^{i\,\odot} \left( 1 - P_{ee} \right) \cos^2\vartheta_{23},
\quad
\Phi_{\nu_{\tau}}^{i}
=
\Phi_{\nu_{e}}^{i\,\odot} \left( 1 - P_{ee} \right) \sin^2\vartheta_{23},
,
\label{eq:solflux}
\end{equation}
where
$\Phi_{\nu_{e}}^{i\,\odot}$ are the fluxes of $\nu_{e}$ produced
by thermonuclear reactions in the center of the Sun,
with $i = pp, \, ^{7}\text{Be}, \, \text{etc.}$,
and $P_{ee}$ is the survival probability of $\nu_{e}$ at the Earth.
In our analysis we use the solar spectra taken from Ref.~\cite{bahcall_web,Bahcall:1987jc,Bahcall:1994cf,Bahcall:1996qv} using the normalizations for the high metallicity model taken from the review in Ref.~\cite{Villante:2020adi}.
We consider $pp$ and $^{7}$Be neutrinos,
which give the main contribution to the
event rates of the experiments under consideration.
For these low-energy solar neutrinos
\begin{equation}
P_{ee}
\simeq
\left( 1 - \frac{1}{2} \, \sin^2 2\vartheta_{12} \right)
\cos^4\vartheta_{13}
+
\sin^4\vartheta_{13}
.
\label{Pee}
\end{equation}
Using
$ \sin^2\vartheta_{12} = 0.318 \pm 0.016 $
and
$ \left. \sin^2\vartheta_{13} \right|_{\text{NO}} = 0.02200^{+0.00069}_{-0.00062} $
or
$ \left. \sin^2\vartheta_{13} \right|_{\text{IO}} = 0.02225^{+0.00064}_{-0.00070} $
obtained in the global fit of Ref.~\cite{deSalas:2020pgw}
(see also the similar results obtained in Refs.~\cite{Esteban:2020cvm,Capozzi:2021fjo})
in the case of normal ordering and inverted ordering 
of the neutrino masses,
we obtain
$ P_{ee} = 0.542 \pm 0.011 $.
The fluxes of $\nu_{\mu}$ and $\nu_{\tau}$
depend on the mixing angle $\vartheta_{23}$,
which is close to maximal mixing ($\vartheta_{23}=\pi/4$)~\cite{deSalas:2020pgw,Esteban:2020cvm,Capozzi:2021fjo}.
For simplicity, we consider
$\sin^2\vartheta_{23}=0.5$,
which implies
$\Phi_{\nu_{\mu}}^{i}=\Phi_{\nu_{\tau}}^{i}$.
Therefore,
we obtain equal constraints on the electromagnetic properties of $\nu_{\mu}$ and $\nu_{\tau}$.

\subsection{The Standard Model \eves cross section}
\label{sub:SMcs}

The Standard Model \eves cross section per Xenon atom is given by
\begin{equation}
\dfrac{d\sigma_{\nu_{\ell}-\text{Xe}}^{\text{SM}}}{d T_{\text{e}}}
(E_{\nu},T_{\text{e}})
=
Z_{\text{eff}}^{\text{Xe}}(T_{e})
\,
\dfrac{G_{\text{F}}^2 m_{e}}{2\pi}
\left[
\left( g_{V}^{\nu_{\ell}} + g_{A}^{\nu_{\ell}} \right)^2
+
\left( g_{V}^{\nu_{\ell}} - g_{A}^{\nu_{\ell}} \right)^2
\left( 1 - \dfrac{T_{e}}{E_{\nu}} \right)^2
-
\left( (g_{V}^{\nu_{\ell}})^2 - (g_{A}^{\nu_{\ell}})^2 \right)
\dfrac{m_{e} T_{e}}{E_{\nu}^2}
\right]
,
\label{eq:ES-SM}
\end{equation}
where
$G_{\text{F}}$ is the Fermi constant,
$m_{e}$ is the electron mass,
$E_{\nu}$ is the neutrino energy, and
$T_e$ is the observable electron recoil energy.
The neutrino-electron couplings $g_{V,A}^{\nu_{\ell}}$
depend on the neutrino flavor:
\begin{align}
\null & \null
g_{V}^{\nu_{e}}
=
2\sin^{2} \vartheta_{W} + 1/2
,
\quad
\null && \null
g_{A}^{\nu_{e}}
=
1/2
,
\label{gnue}
\\
\null & \null
g_{V}^{\nu_{\mu,\tau}}
=
2\sin^{2} \vartheta_{W} - 1/2
,
\quad
\null && \null
g_{A}^{\nu_{\mu,\tau}}
=
- 1/2
,
\label{gnumu}
\end{align}
with $\sin^{2} \vartheta_{W}=0.23863\pm0.00005$~\cite{ParticleDataGroup:2022pth}.
The coefficient $Z_{\text{eff}}^{\text{Xe}}(T_{e})$
quantifies the effective number of electrons which can be ionized at $T_e$~\cite{Fayans:2000ns}.
We calculate $Z_{\text{eff}}^{\text{Xe}}(T_{e})$ using Table~II
of Ref.~\cite{AtzoriCorona:2022jeb}.

\subsection{Neutrino magnetic moments}
\label{sub:MMcs}

Neutrino magnetic moments are predicted by many theories
with massive neutrinos beyond the Standard Model (BSM)
and have been constrained by many observations
(see, e.g., Refs.~\cite{Giunti:2014ixa,Giunti:2015gga,Giunti:2022aea,ParticleDataGroup:2022pth}).
The most stringent experimental limit
$ |\mu_{\nu_e}| < 2.9 \times 10^{-11} \, \mu_{B} $ at 90\% C.L.,
where $\mu_{B}$ is the Bohr magneton,
was obtained in the GEMMA experiment~\cite{Beda:2012zz}
through the \eves of reactor $\bar\nu_{e}$.
This bound is more than eight orders of magnitude larger than the
prediction of neutrino magnetic moments
in the minimal extension of the SM with right handed neutrinos
and Dirac neutrino masses~\cite{Fujikawa:1980yx,Pal:1981rm,Shrock:1982sc}.
However,
in more elaborate models
(see, e.g., the review in Ref.~\cite{Giunti:2014ixa}),
the neutrino magnetic moments can be larger and can be probed in current experiments.

The contribution of the neutrino magnetic moments to \eves
is incoherent with the Standard Model contribution
because the neutrino magnetic moment interaction flips the helicity
of ultrarelativistic neutrinos,
whereas the Standard Model weak interaction is helicity-conserving.
Therefore,
for each flavor neutrino $\nu_{\ell}$
the total cross section is given by
the sum of the Standard Model cross section~\eqref{eq:ES-SM}
and the magnetic moment cross section
\begin{equation}
\dfrac{d\sigma_{\nu_{\ell}\text{-}\text{Xe}}^{\text{MM}}}{d T_\text{e}}
(E_{\nu},T_\text{e})
=
Z_{\text{eff}}^{\text{Xe}}(T_{\text{e}}) \dfrac{ \pi \alpha^2 }{ m_{e}^2 }
\left( \dfrac{1}{T_\text{e}} - \dfrac{1}{E_{\nu}} \right)
\left| \dfrac{\mu_{\nu_{\ell}}}{\mu_{\text{B}}} \right|^2
,
\label{ES-MM}
\end{equation}
where $\alpha$ is the fine-structure constant.
For the low values of
$T_\text{e}$
in the experiments under consideration
($ T_\text{e} \lesssim 30 \, \text{keV} $),
the cross section \eqref{ES-MM}
is approximately proportional to $T_\text{e}^{-1}$.
Therefore, the effects of the neutrino magnetic moments
are probed by the observation of an event excess near the $T_\text{e}$ threshold.

\subsection{Neutrino electric charges}
\label{sub:ECcs}

In the Standard Model neutrinos are neutral,
but in BSM theories they can have small electric charges,
often called
``millicharges''
(see, e.g., the review in Ref.~\cite{Giunti:2014ixa}),
which can be probed in neutrino scattering experiments.
In general, the three flavor neutrinos can have the millicharges
$q_{\nu_{e}}$, $q_{\nu_{\mu}}$, and $q_{\nu_{\tau}}$,
and there can be also the three transition electric charges
$q_{\nu_{e\mu}}$, $q_{\nu_{e\tau}}$, and $q_{\nu_{\mu\tau}}$.

Since the electric charge interaction conserves the helicity
of ultrarelativistic neutrinos and the millicharges of the flavor neutrinos
conserve the neutrino flavor,
they contribute coherently
with the Standard Model interaction,
which is helicity and flavor conserving.
On the other hand,
since the transition electric charges induce a change of flavor
they contribute incoherently
with the Standard Model interaction.
Therefore,
the total \eves cross section is given by
\begin{equation}
\dfrac{d\sigma_{\nu_{\ell}-\text{Xe}}^{\text{SM+EC}}}{d T_{\text{e}}}
=
\left(
\dfrac{d\sigma_{\nu_{\ell}-\text{Xe}}^{\text{SM+EC}}}{d T_{\text{e}}}
\right)_{q_{\nu_{\ell}}}
+
\sum_{\ell'\neq\ell}
\left(
\dfrac{d\sigma_{\nu_{\ell}-\text{Xe}}^{\text{EC}}}{d T_{\text{e}}}
\right)_{q_{\nu_{\ell\ell'}}}
,
\label{ES-EC}
\end{equation}
where
$ \left(
d\sigma_{\nu_{\ell}-\text{Xe}}^{\text{SM+EC}} / d T_{\text{e}}
\right)_{q_{\nu_{\ell}}}$
is given by Eq.~\eqref{eq:ES-SM} with
\begin{equation}
g_{V}^{\nu_{\ell}}
\to
g_{V}^{\nu_{\ell}}
-
\dfrac{ \sqrt{2} \pi \alpha }{ G_{\text{F}} m_{e} T_{e} }
\, q_{\nu_{\ell}}
,
\label{eq:gV-EC}
\end{equation}
and
\begin{equation}
\left(
\dfrac{d\sigma_{\nu_{\ell}-\text{Xe}}^{\text{EC}}}{d T_{\text{e}}}
\right)_{q_{\nu_{\ell\ell'}}}
=
Z_{\text{eff}}^{\text{Xe}}(T_{e})
\,
\dfrac{\pi \alpha^2}{m_{e} T_{\text{e}}^2}
\left[
1
+
\left( 1 - \dfrac{T_{\text{e}}}{E_{\nu}} \right)^2
-
\dfrac{m_{e} T_{\text{e}}}{E_{\nu}^2}
\right]
|q_{\nu_{\ell\ell'}}|^2
,
\label{ES-EC-tr}
\end{equation}
for $\ell'\neq\ell$.
Hence, \eves gives full information on the charges
$q_{\nu_{\ell}}$
of the flavor neutrinos,
including their sign,
whereas only the absolute values of the transition electric charges
$q_{\nu_{\ell\ell'}}$
can be probed.
Note also that the effects of the electric charges are enhanced
at small values of $T_{e}$,
leading to the possibility to probe very small electric charges
in low-threshold experiments.

\subsection{Neutrino charge radii}
\label{sub:CRcs}

Even if neutrinos are neutral,
they can have charge radii.
Indeed, even in the Standard Model the massless and neutral flavor neutrinos have tiny charge radii induced by radiative corrections,
which are given by~\cite{Bernabeu:2000hf,Bernabeu:2002nw,Bernabeu:2002pd}
(with the definition of the charge radii in Refs.~\cite{Cadeddu:2018dux,Cadeddu:2019eta})
\begin{equation}
\langle{r}_{\nu_{\ell}}^2\rangle_{\text{SM}}
=
-
\frac{G_{\text{F}}}{2\sqrt{2}\pi^2}
\left[
3-2\ln\left(\frac{m_{\ell}^2}{m^2_{W}}\right)
\right]
,
\label{G050}
\end{equation}
where $m_{W}$ and $m_{\ell}$ are, respectively,
the $W$ boson and charged lepton masses
($\ell = e, \mu, \tau$).
Numerically, we have
\begin{align}
\null & \null
\langle{r}_{\nu_{e}}^2\rangle_{\text{SM}}
=
- 0.83 \times 10^{-32} \, \text{cm}^2
,
\label{reSM}
\\
\null & \null
\langle{r}_{\nu_{\mu}}^2\rangle_{\text{SM}}
=
- 0.48 \times 10^{-32} \, \text{cm}^2
,
\label{rmSM}
\\
\null & \null
\langle{r}_{\nu_{\tau}}^2\rangle_{\text{SM}}
=
- 0.30 \times 10^{-32} \, \text{cm}^2
.
\label{rtSM}
\end{align}
These diagonal charge radii
of the flavor neutrinos
are the only charge radii that exist in the Standard Model,
where neutrino flavor is conserved.

In BSM theories neutrinos can have also the transition charge radii
$\langle{r}_{\nu_{e\mu}}^2\rangle$,
$\langle{r}_{\nu_{e\tau}}^2\rangle$, and
$\langle{r}_{\nu_{\mu\tau}}^2\rangle$
which induce flavor transitions in scattering processes.
We consider the general case with both diagonal and transition charge radii.
As for the electric charges discussed in Subsection~\ref{sub:ECcs},
the diagonal charge radii contribute to the \eves cross section coherently
with the Standard Model interaction,
because both conserve
the helicity
of ultrarelativistic neutrinos and neutrino flavors,
whereas the transition charge radii contribute incoherently.
Therefore,
the total \eves cross section is given by
\begin{equation}
\dfrac{d\sigma_{\nu_{\ell}-\text{Xe}}^{\text{SM+CR}}}{d T_{\text{e}}}
=
\left(
\dfrac{d\sigma_{\nu_{\ell}-\text{Xe}}^{\text{SM+CR}}}{d T_{\text{e}}}
\right)_{\langle{r}_{\nu_{\ell}}^2\rangle}
+
\sum_{\ell'\neq\ell}
\left(
\dfrac{d\sigma_{\nu_{\ell}-\text{Xe}}^{\text{CR}}}{d T_{\text{e}}}
\right)_{\langle{r}_{\nu_{\ell\ell'}}^2\rangle}
,
\label{ES-CR}
\end{equation}
where
$ \left(
d\sigma_{\nu_{\ell}-\text{Xe}}^{\text{SM+CR}} / d T_{\text{e}}
\right)_{\langle{r}_{\nu_{\ell}}^2\rangle}$
is given by Eq.~\eqref{eq:ES-SM} with
\begin{equation}
g_{V}^{\nu_{\ell}}
\to
g_{V}^{\nu_{\ell}}
+
\dfrac{ \sqrt{2} \pi \alpha }{ 3 G_{\text{F}} }
\, \langle{r}_{\nu_{\ell\ell'}}^2\rangle
,
\label{gV-CR}
\end{equation}
and
\begin{equation}
\left(
\dfrac{d\sigma_{\nu_{\ell}-\text{Xe}}^{\text{CR}}}{d T_{\text{e}}}
\right)_{\langle{r}_{\nu_{\ell\ell'}}^2\rangle}
=
Z_{\text{eff}}^{\mathcal{A}}(T_{e})
\,
\dfrac{\pi \alpha^2 m_{e}}{9}
\left[
1
+
\left( 1 - \dfrac{T_{e}}{E_{\nu}} \right)^2
-
\dfrac{m_{e} T_{e}}{E_{\nu}^2}
\right]
|\langle{r}_{\nu_{\ell\ell'}}^2\rangle|^2
,
\label{ES-CR-tr}
\end{equation}
for $\ell'\neq\ell$.
As for the electric charges discussed in Subsection~\ref{sub:ECcs},
\eves gives full information on the diagonal charge radii
$\langle{r}^2_{\nu_{\ell}}\rangle$
of the flavor neutrinos,
including their sign,
and only information on the absolute values of the transition electric charges
$\langle{r}_{\nu_{\ell\ell'}}^2\rangle$.

\section{Data analysis}
\label{sec:analysis}

\subsection{Current experiments}
In the analyses presented in this paper we include data from LZ~\cite{LZ:2022lsv}, XENONnT~\cite{XENON:2022ltv} and PandaX-4T~\cite{PandaX:2022ood}. In this section we discuss the details of each analysis. For an experiment $X$ the overall predicted number of events in a given energy-bin $k$ is given by 
\begin{equation}
    R^X_k = R^{E\nu ES}_k + \sum_i R^i_k\,,
\label{eq:pred_n_evs}
\end{equation}
where $R^{E\nu ES}_k$ is the contribution from solar neutrinos which elastically scatter on electrons, while $R^i_k$ are the remaining background components. We have extracted the contributions $R^i_k$ for the different experiments from Refs.~\cite{LZ:2022ysc,XENON:2022ltv,PandaX:2022ood} and $R^{E\nu ES}_k$ is obtained from 
\begin{equation}
   R^{E\nu ES}_k = N ~\int_{T_e^k}^{T_e^{k+1}}dT_e ~\int_0^\infty dT_e'~ R(T_e,T_e')~ A(T_e') 
   \sum_{i=pp,^{7}\text{Be}}
   \int_{E_{\nu}^{\text{min}}}^{E_{\nu,i}^{\text{max}}} dE_\nu ~\sum_\ell ~\Phi_{\nu_\ell}^{i}(E_\nu)~ \frac{d\sigma_{\nu_\ell}}{dT_e'}\,.
\label{eq:r_eves}
\end{equation}
In this expression $T_e$ and $T_e'$ are the reconstructed and true electron recoil energies, while $E_\nu$ is the neutrino energy. The electron-neutrino cross section for a given neutrino flavor $\nu_\ell$ is given by $d\sigma_{\nu_\ell} / dT_e'$ and $\Phi_{\nu_\ell}^{i}(E_\nu)$ are the solar neutrino fluxes from Eq.~\eqref{eq:solflux}. The minimal neutrino energy to produce an electron recoil of $T_e'$ is given by $E_\nu^{\text{min}} = (T_e' + \sqrt{2m_eT_e'+T_e'^2})/2$, while the maximal neutrino energy $E_{\nu,i}^{\text{max}}$ depends on the production process, indicated with the index $i$.  Next, $R(T_e,T_e')$ and $A(T_e')$ are the detector resolution and efficiency and are different for all experiments. Finally, $N$ is a normalization constant which takes into account the exposure and detector volume. Also this quantity is different for each experiment. 

For the analysis of the experiments under consideration we use the detector efficiencies from Refs.~\cite{LZ:2023poo,XENON:2022ltv,PandaX:2022ood}. For the energy resolution at LZ we use the same function that has been used in Refs.~\cite{AtzoriCorona:2022jeb,A:2022acy}. In the case of XENONnT we use the resolution function of Ref.~\cite{XENON:2020rca} and for PandaX-4T we use the one from Ref.~\cite{PandaX:2022ood}.

The predicted number of events in Eq.~\eqref{eq:pred_n_evs} has to be compared with the data $D^X$ accumulated in each experiment. We use the data presented in Fig.~6 of Ref.~\cite{LZ:2022lsv} for LZ\footnote{We do not use the timing data from Ref.~\cite{LZ:2023poo}, which could result only in a 5\% improvement in the bounds.} and the data in Fig.~3 of Ref.~\cite{PandaX:2022ood} for PandaX-4T. For XENONnT we use the data from Fig.~4 (5) of Ref.~\cite{XENON:2022ltv} for recoil energies above (below) 30~keV. 

Due to the low statistics in some of the bins, for LZ and PandaX-4T we use the Poissonian least-squares function
\begin{equation}
    \chi^2_X = \min_{\vec{\alpha},\vec{\beta}} \left\{2\left(\sum_k R^X_k - D^X_k + D^X_k~\log D^X_k/R^X_k\right) + \sum_i (\alpha_i/\sigma_{\alpha_i})^2 + \sum_i (\beta_i/\sigma_{\beta_i})^2\right\}\,,
\end{equation}
where $\alpha_i$ are normalization constants multiplied to each single background component in Eq.~\eqref{eq:pred_n_evs}.
For LZ, the uncertainties $\sigma_{\alpha_i}$ are extracted from Ref.~\cite{LZ:2022ysc}.
For PandaX-4T, they are taken from Ref.~\cite{PandaX:2022ood}.
Note that some of them are left to vary freely in the analysis.
Also included are uncertainty coefficients  $\beta_i$ of the solar neutrino fluxes, with uncertainties $\sigma_{\beta_i}$ taken from Ref.~\cite{Villante:2020adi}. 

In the case of XENONnT data, we use instead
\begin{equation}
    \chi^2_{\text{XENONnT}} = \min_{\vec{\alpha},\vec{\beta}} \left\{\sum_k\left(\frac{R^{\text{XENONnT}}_k - D^{\text{XENONnT}}_k}{\sigma_k}\right)^2 + \sum_i (\alpha_i/\sigma_{\alpha_i})^2 + \sum_i (\beta_i/\sigma_{\beta_i})^2\right\}\,,
\end{equation}
where the uncertainties in each bin $\sigma_k$ are extracted from Ref.~\cite{XENON:2022ltv}. The remaining components are equivalent to the corresponding ones for LZ and PandaX-4T. 

We also perform a combined analysis of all three experiments. In this case we correlate the uncertainties regarding the neutrino flux among the experiments. In addition, several of the background components are common to at least two of the three experiments. In these cases we also correlate the normalizations of the background components. While such a combined analysis has not been performed yet, we have noticed that this correlated analysis produces only slightly tighter bounds than simply summing up the individual $\chi^2$.

\subsection{DARWIN sensitivity}

We also compute the sensitivity to electromagnetic neutrino properties for the future experiment DARWIN. The calculation of the event rate at DARWIN is basically the same as for the current experiments, given in Eqs.~\eqref{eq:pred_n_evs} and~\eqref{eq:r_eves}, but we also include the contributions from solar N, O and $pep$ neutrinos. Note, however, that their contribution is mostly negligible in comparison with some of the background contributions, as shown in Fig.~1 of Ref.~\cite{DARWIN:2020bnc}. We include them, nevertheless, since we use the full spectrum as shown in Ref.~\cite{DARWIN:2020bnc}. The energy dependence of the neutrino oscillation probability is taken into account for these higher-energy neutrinos. 

For the individual background components we use the spectra given in Ref.~\cite{DARWIN:2020bnc}, which need to be normalized to the considered exposure. Due to lack of more detailed information, we use the same resolution function and detector efficiency as for XENONnT. We assume the efficiency to remain constant for $T_e>T_{e,\text{max}}^{\text{XENONnT}}$. With these assumptions, we are able to reproduce the E$\nu$ES spectra for all five neutrino species in Fig.~1 of Ref.~\cite{DARWIN:2020bnc}, which validates our choice of efficiency and resolution. As was done in Ref.~\cite{DARWIN:2020bnc}, we consider different scenarios for the DARWIN sensitivity. First we assume an exposure of 30~ty. Next, we assume an exposure of 300~ty. Finally, we assume an exposure of 300~ty again, but also that the $^{136}$Xe background component is depleted to 1\%. 

When generating the mock data, always compatible with the Standard Model, we use 51 logarithmically spaced bins between 1 and 1500~keV recoil energy. Note that the spectrum at higher energies is not sensitive to any BSM effect considered in this paper, because the E$\nu$ES rate is much smaller than some of the background rates. We still use the full spectrum, since the inclusion of events at high energies can help to control the effect of background uncertainties.

\begin{figure}[t]
\begin{subfigure}{0.48\textwidth}
    \includegraphics[width=\textwidth]{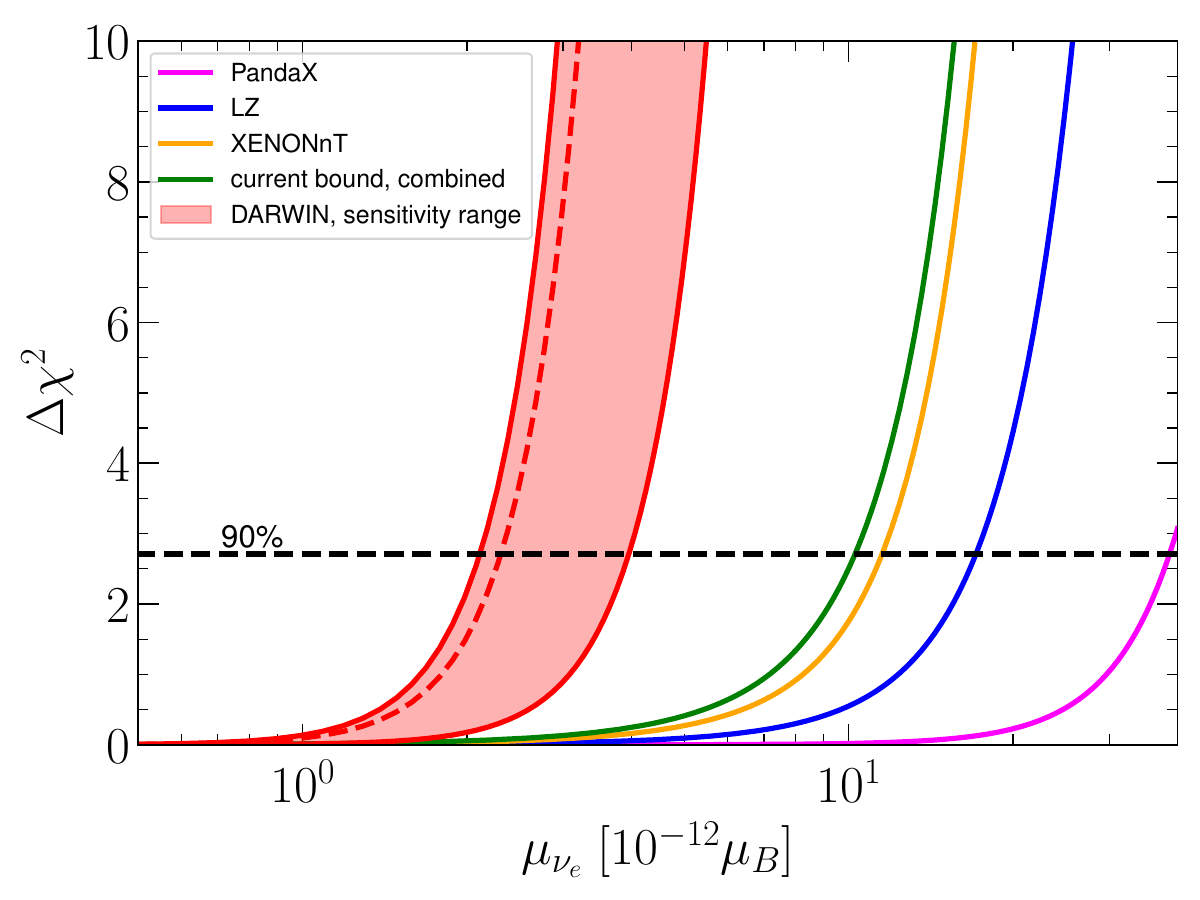}
\end{subfigure}
\hfill
\begin{subfigure}{0.48\textwidth}
    \includegraphics[width=\textwidth]{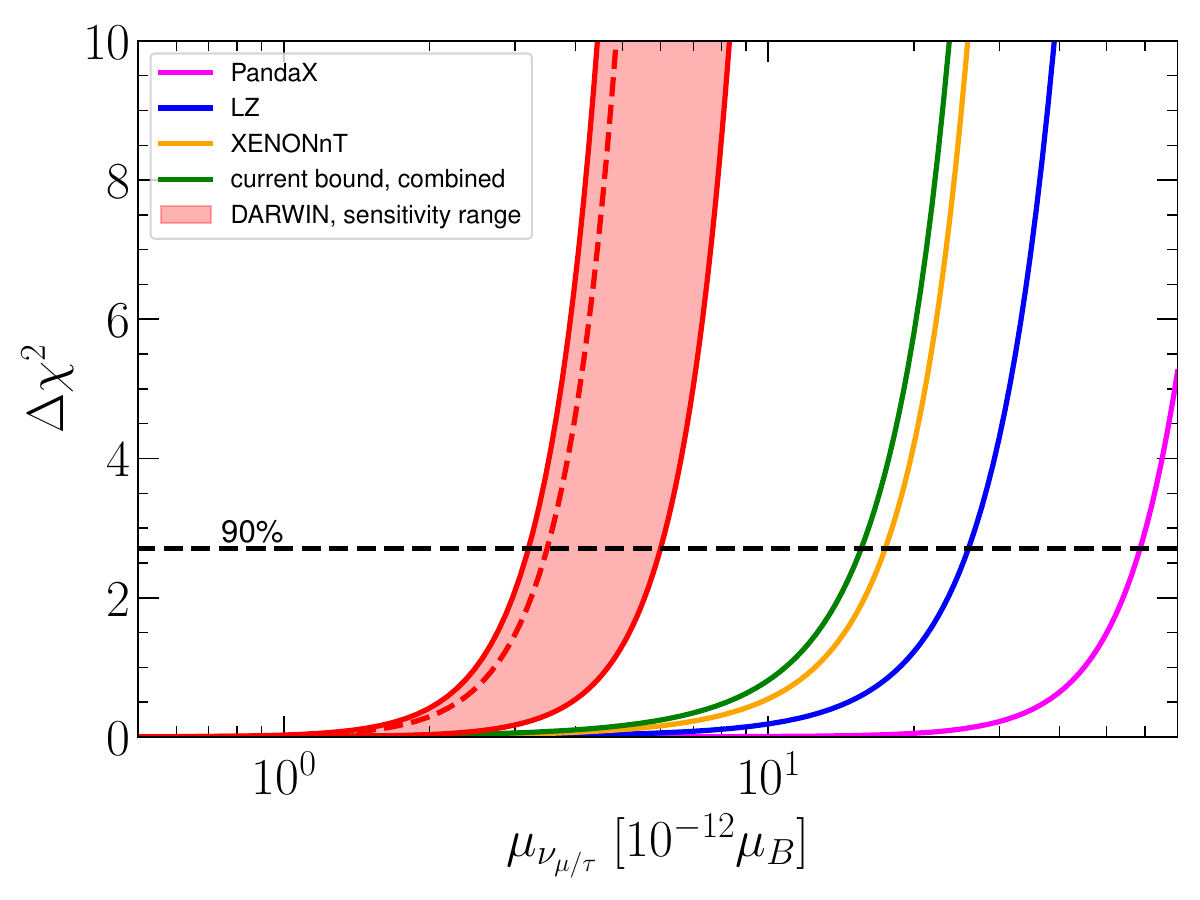}
    %caption
    %label
\end{subfigure}
\hfill
\begin{subfigure}{0.48\textwidth}
    \includegraphics[width=\textwidth]{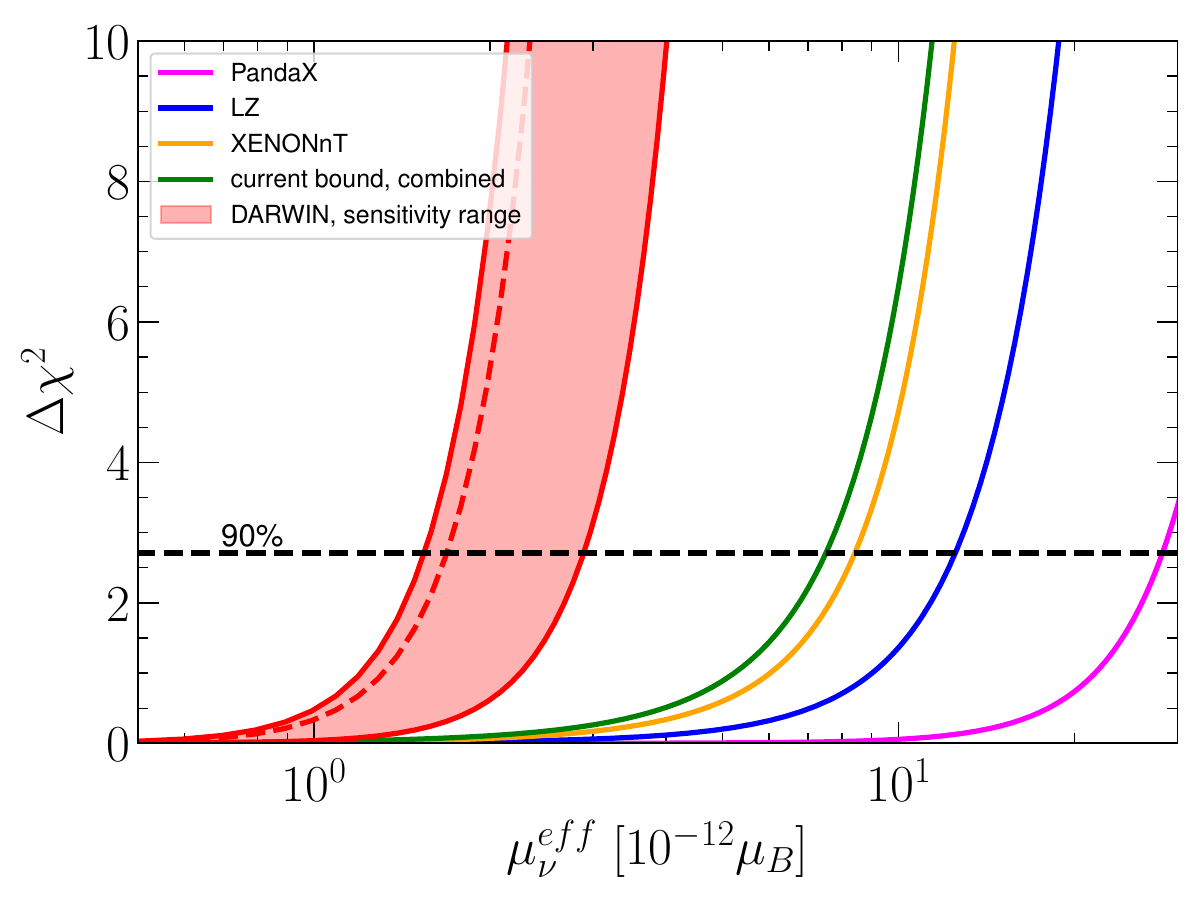}
    %caption
    %label
\end{subfigure}
\hfill
\caption{$\Delta\chi^2$ profiles for the different neutrino magnetic moments obtained from the analyses of the data of PandaX (magenta), LZ (blue), XENONnT (orange) and from the combination of all experiments (green). Also shown is the sensitivity range for DARWIN (red shaded), where the worst (best) case scenario corresponds to the analysis with 30~ty (300~ty) exposure without (with) depleted $^{136}$Xe background. The dashed line corresponds to 300~ty exposure without depleted $^{136}$Xe background.}
\label{fig:magnetic_moment}
\end{figure}

\section{Results}
\label{sec:res}

In this section we present the bounds that can be obtained from current data and the sensitivity at DARWIN to neutrino electromagnetic properties.

\subsection{Neutrino magnetic moments}
\label{sec:magn_mom}

We first discuss neutrino magnetic moments. Our results are presented in Fig.~\ref{fig:magnetic_moment} and Table~\ref{tab:magn_mom}. In the first column of Table~\ref{tab:magn_mom} we present the bounds on the magnetic moment $\mu_{\nu_e}$. In the second we present the results for $\mu_{\nu_\mu}$ and $\mu_{\nu_\tau}$, which are the same in our analysis. Slightly different bounds for the two moments have been found, e.g., in Refs.~\cite{AtzoriCorona:2022jeb,A:2022acy},
where non maximal atmospheric mixing ($\sin^2\theta_{23}\neq0.5$) was assumed to calculate the neutrino oscillation probability. Since we chose maximal atmospheric mixing,
the expected number of events is the same for a non-zero $\mu_{\nu_\mu}$ or $\mu_{\nu_\tau}$, hence producing the same bound. Finally, we consider the case of a single effective magnetic moment, $\mu_\nu^{eff} = \mu_{\nu_e} = \mu_{\nu_\mu} = \mu_{\nu_\tau}$.

Our analysis of PandaX-4T updates the analysis of Ref.~\cite{PandaX-II:2020udv}, where the Panda collaboration analyzed the data from a smaller version of the current detector. In any case, due to the larger background rates observed in this experiment the bounds are always weaker than those obtained from LZ or XENONnT. It should be noted that the LZ bound obtained in our analysis is a bit weaker than those obtained in Refs.~\cite{AtzoriCorona:2022jeb,A:2022acy}. This is due to the more conservative use of systematic uncertainties in our analysis, since we treat each background component individually with an individual nuisance parameter. Also our XENONnT bound is slightly weaker than that from the collaboration~\cite{XENON:2022ltv} and the one from Ref.~\cite{Khan:2022bel}. This is not worrying since we used a different statistical method and since our bound lies in any case within the XENONnT sensitivity band, see Ref.~\cite{XENON:2022ltv}. Also in the case of XENONnT, we are more conservative in the treatment of systematic uncertainties than other phenomenological analyses~\cite{A:2022acy,Khan:2022bel}.

The bounds obtained from the combined analysis, which at 90\% confidence level read
\begin{eqnarray}
    |\mu_{\nu_e}| &<& 10.3\times10^{-12}~\mu_B\,, \\
    |\mu_{\nu_{\mu/\tau}}| &<& 15.6\times10^{-12}~\mu_B\,, \\
    |\mu_{\nu}^{eff}| &<& 7.5\times10^{-12}~\mu_B\,,
\end{eqnarray}
are among the strongest laboratory bounds on neutrino magnetic moments~\cite{Giunti:2014ixa,Giunti:2015gga,Giunti:2022aea,ParticleDataGroup:2022pth}.
They are about one or two orders of magnitude stronger than bounds from COHERENT or Dresden-II data~\cite{AtzoriCorona:2022qrf,Coloma:2022avw,Liao:2022hno,Khan:2022akj,DeRomeri:2022twg}, or Borexino data~\cite{Coloma:2022umy}. They are also stronger than older bounds reported in the literature~\cite{MUNU:2005xnz,TEXONO:2006xds,Beda:2012zz,Ahrens:1990fp,Allen:1992qe,LSND:2001akn,DONUT:2001zvi,Canas:2015yoa},
as the best bound on $\mu_{\nu_e}$ obtained in the GEMMA experiment
($ |\mu_{\nu_e}| < 2.9 \times 10^{-11} \, \mu_{B} $ at 90\% C.L.)~\cite{Beda:2012zz},
the best bound on $\mu_{\nu_\mu}$ obtained in the LSND experiment
($ |\mu_{\nu_\mu}| < 6.8 \times 10^{-10} \, \mu_{B} $ at 90\% C.L.)~\cite{LSND:2001akn},
and
the best bound on $\mu_{\nu_\tau}$ obtained in the DONUT experiment
($ |\mu_{\nu_\tau}| < 3.9 \times 10^{-7} \, \mu_{B} $ at 90\% C.L.)~\cite{DONUT:2001zvi}.

As also shown in Fig.~\ref{fig:magnetic_moment} and Table~\ref{tab:magn_mom}, these bounds can be further improved by about a factor of 2--5 by the DARWIN experiment (see also the sensitivity analysis in Ref.~\cite{AristizabalSierra:2020zod}) which would make dark matter direct detection experiments competitive with astrophysical observations,
which constrain the neutrino magnetic moments below a few
$ 10^{-12} \, \mu_{B} $~\cite{Ayala:1998qz,Viaux:2013lha,Corsico:2014mpa,Capozzi:2020cbu,Mori:2020qqd}. 

\begin{table}[t]
\centering
\begin{tabular}{|c||c|c|c|}
\hline
Experiment & $|\mu_{\nu_e}| ~[10^{-12}\mu_B]$  & $|\mu_{\nu_{\mu/\tau}}| ~[10^{-12}\mu_B]$ & $|\mu_{\nu}^{eff}| ~[10^{-12}\mu_B]$\\
\hline
PandaX-4T & $<38.7$ & $<58.6$ & $<28.3$\\
\hline
LZ & $<17.1$ & $<25.9$ & $<12.5$  \\
\hline
XENONnT & $<11.5$ & $<17.5$ & $<8.4$\\
\hline
combined & $<10.3$ & $<15.6$ & $<7.5$\\
\hline
\hline
DARWIN 30~ty & $<4.0$ & $<6.0$ & $<2.9$\\
\hline
DARWIN 300~ty & $<2.3$ & $<3.5$ & $<1.7$\\
\hline
DARWIN 300~ty depl. & $<2.1$ & $<3.2$ & $<1.5$\\
\hline
\end{tabular}
\caption{The 90\% bounds ($\Delta\chi^2=2.71$) that can be obtained on the different neutrino magnetic moments.}
\label{tab:magn_mom}
\end{table}

\subsection{Neutrino millicharge}
\label{sec:nu_millicharge}

\begin{figure}
\begin{subfigure}{0.48\textwidth}
    \includegraphics[width=\textwidth]{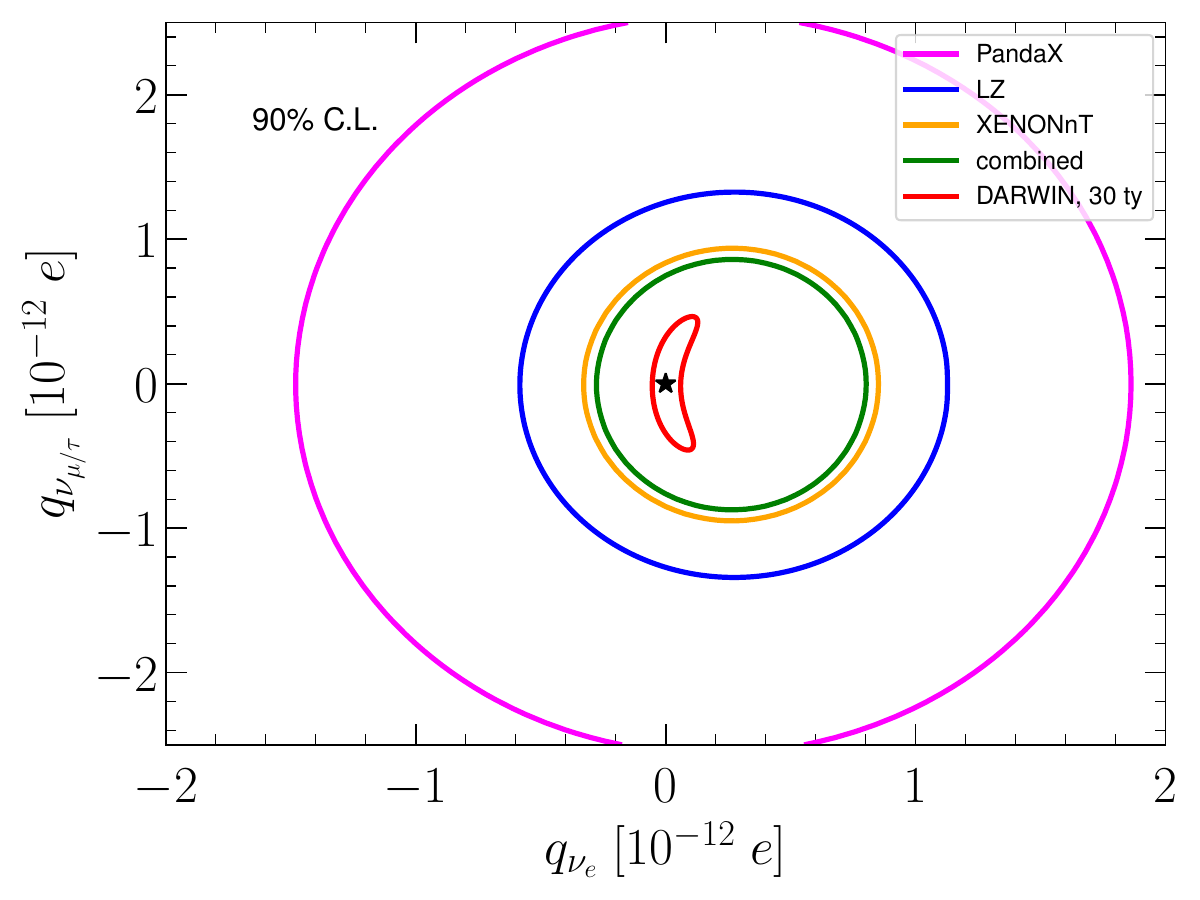}
\end{subfigure}
\hfill
\begin{subfigure}{0.48\textwidth}
    \includegraphics[width=\textwidth]{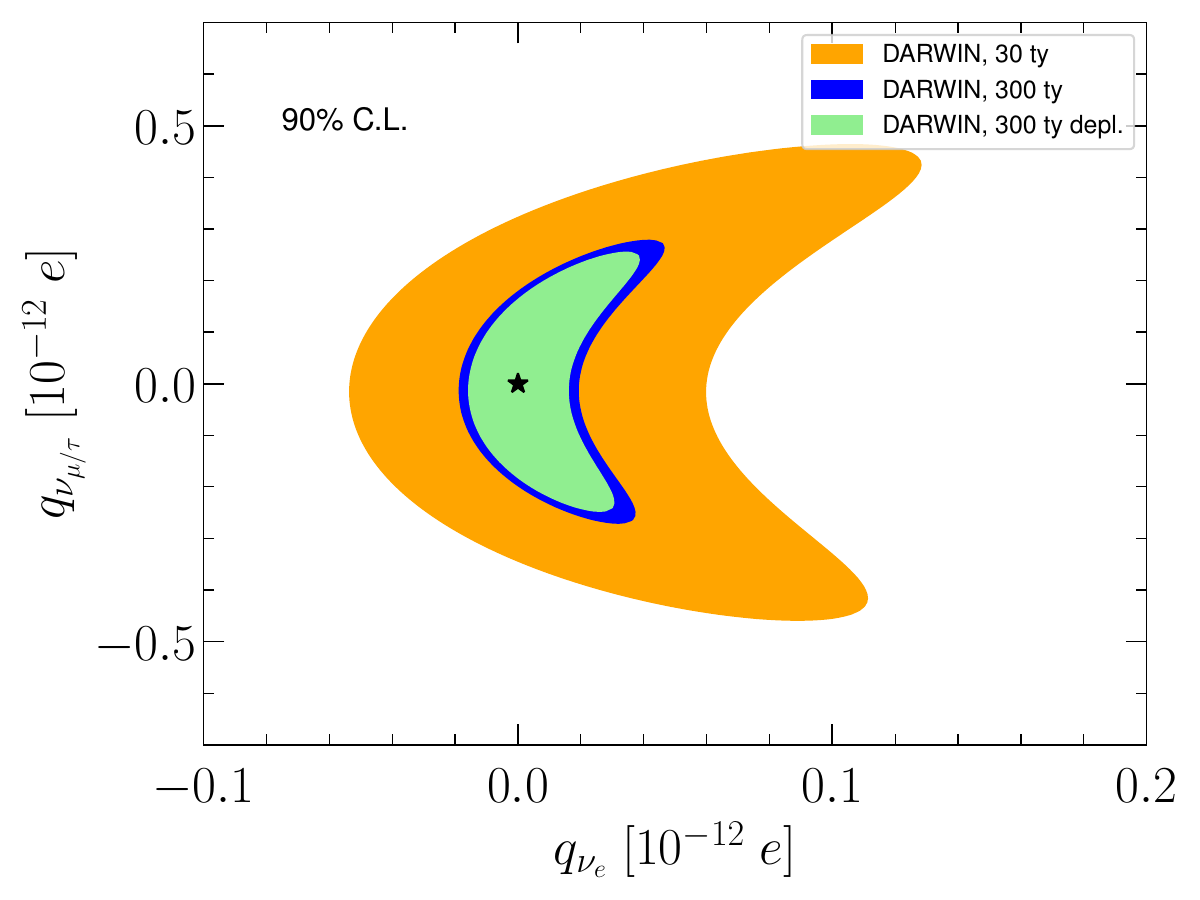}
    %caption
    %label
\end{subfigure}
\hfill
\caption{Left: The 90\% C.L. (2 d.o.f.) allowed regions in the $q_{\nu_{e}}-q_{\nu_{\mu}}$-plane from PandaX (magenta), LZ (blue), XENONnT (orange) and from the combination of all experiments (green) in comparison with the DARWIN sensitivity assuming 30~ty of exposure. Right: The 90\% C.L. (2 d.o.f.) expected sensitivity of DARWIN for different exposures and background assumptions. The star denotes the SM value.}
\label{fig:nu_millicharge_2D}
\end{figure}

\begin{figure}[t]
\begin{subfigure}{0.48\textwidth}
    \includegraphics[width=\textwidth]{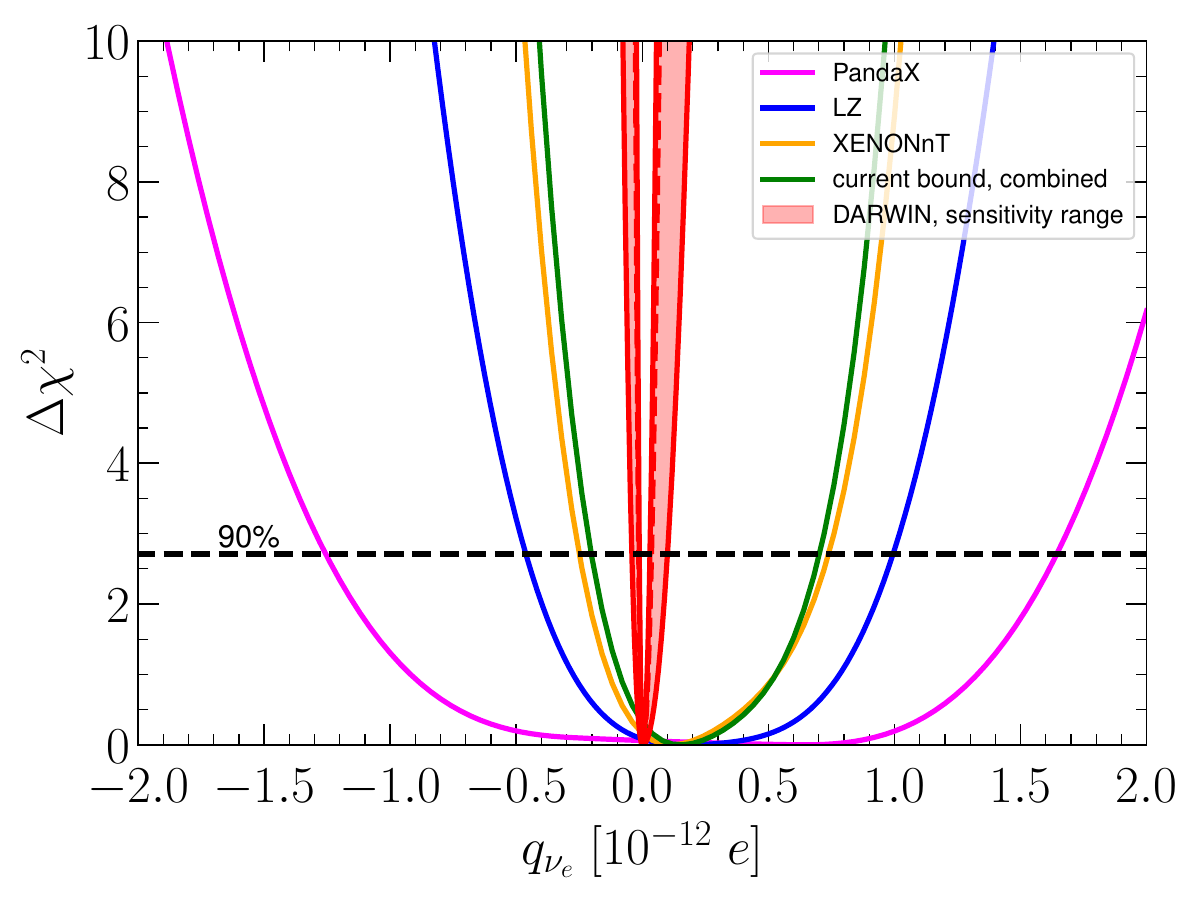}
\end{subfigure}
\hfill
\begin{subfigure}{0.48\textwidth}
    \includegraphics[width=\textwidth]{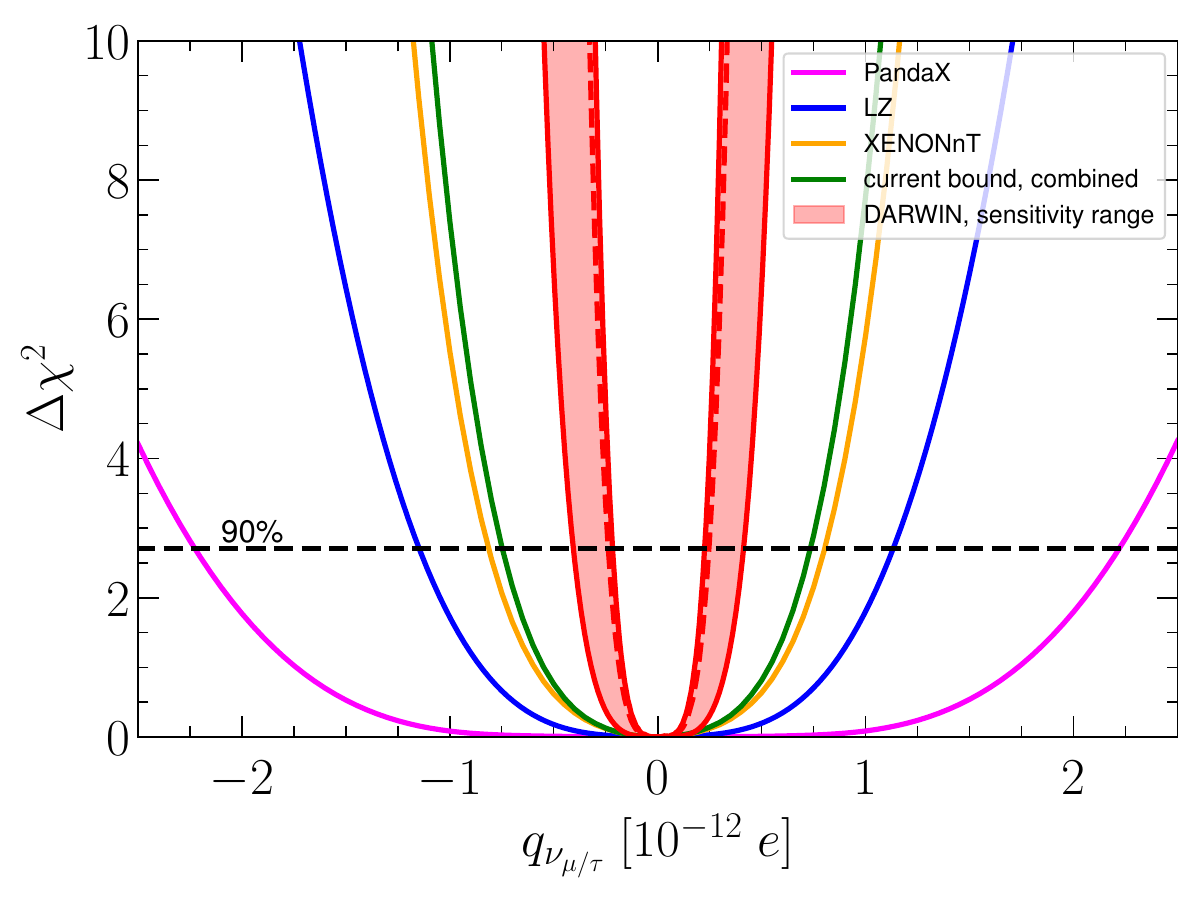}
    %caption
    %label
\end{subfigure}
\hfill
\begin{subfigure}{0.48\textwidth}
    \includegraphics[width=\textwidth]{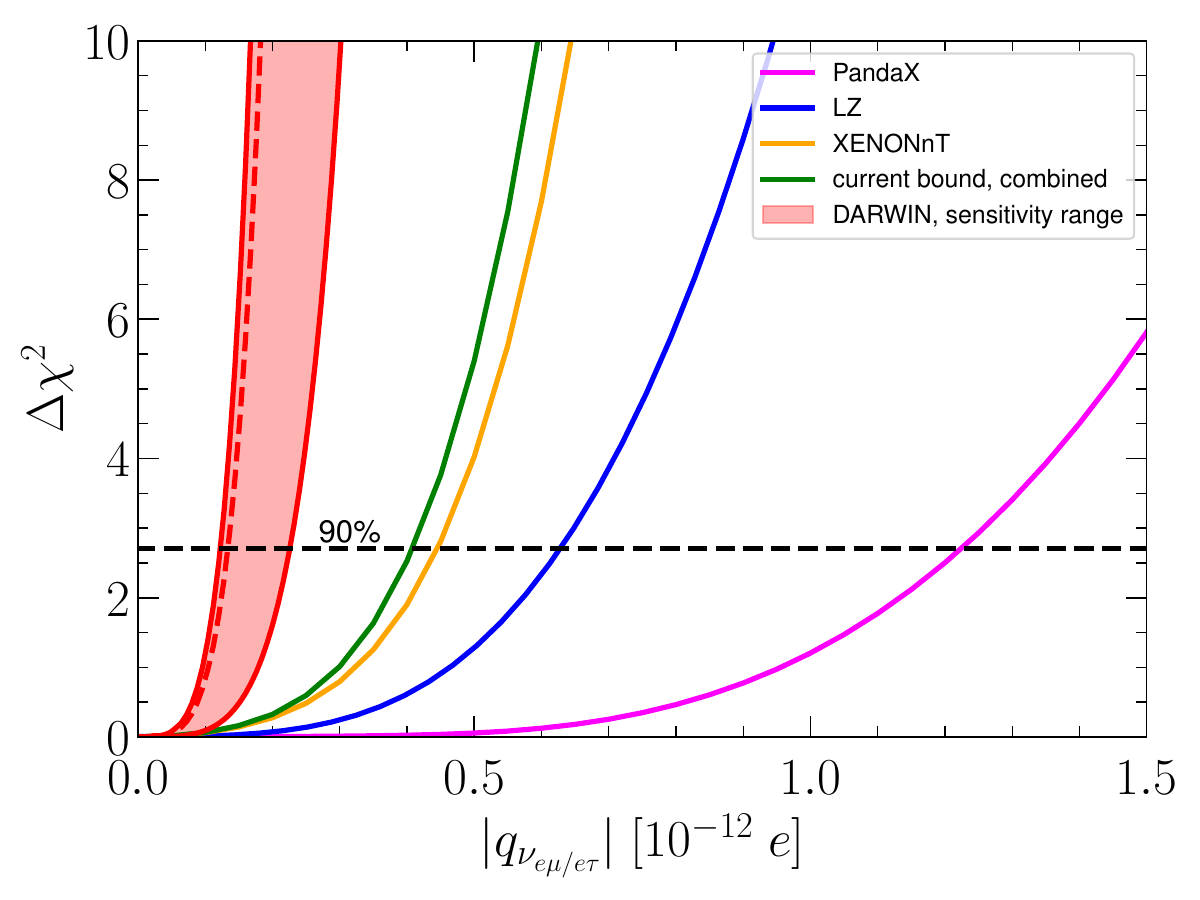}
    %caption
    %label
\end{subfigure}
\hfill
\hfill
\begin{subfigure}{0.48\textwidth}
    \includegraphics[width=\textwidth]{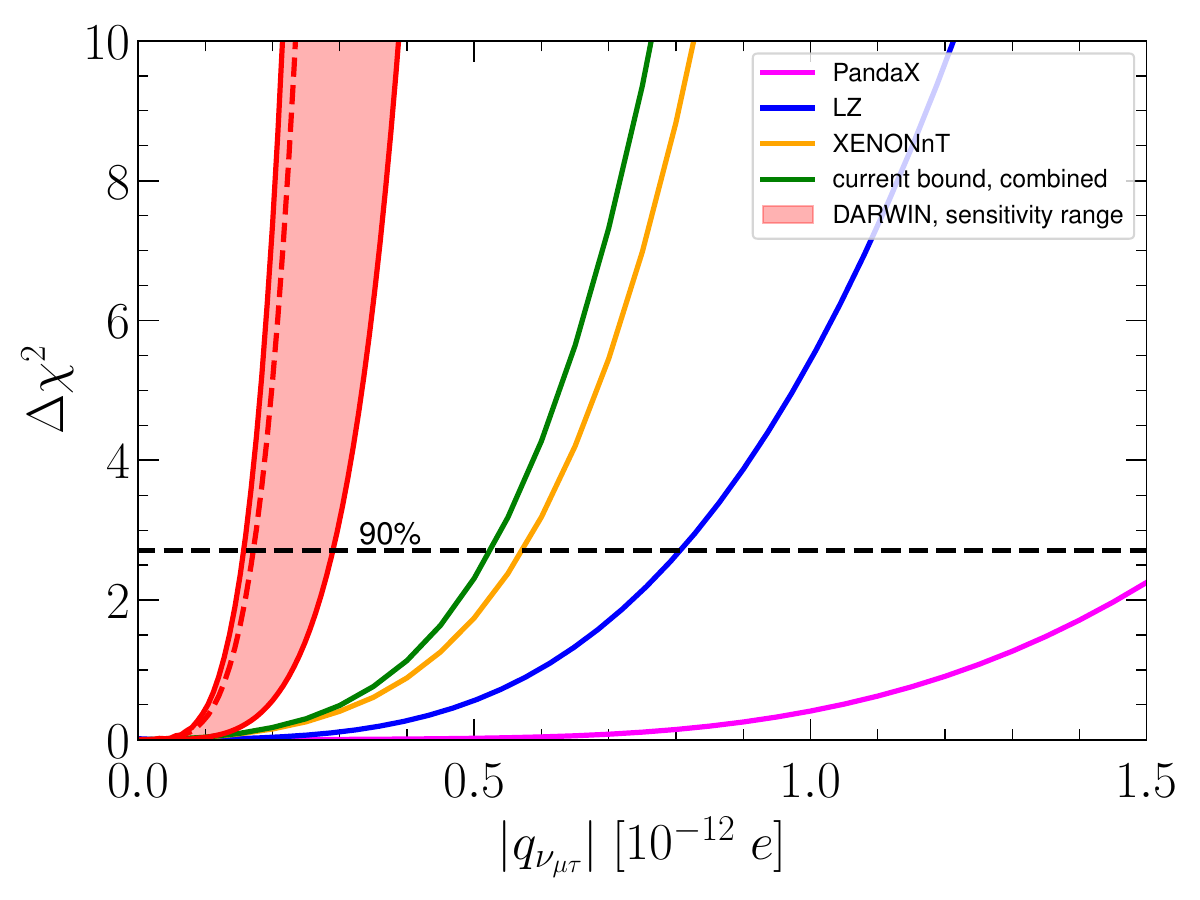}
    %caption
    %label
\end{subfigure}
\hfill
\caption{$\Delta\chi^2$ profiles for the different neutrino millicharges obtained from the analysis of data of PandaX (magenta), LZ (blue), XENONnT (orange) and from the combination of all experiments (green). Also shown is the sensitivity range for DARWIN (red shaded), where the worst (best) case scenario corresponds to the analysis with 30~ty (300~ty) exposure without (with) depleted $^{136}$Xe background. The dashed line corresponds to 300~ty exposure without depleted $^{136}$Xe background.}
\label{fig:nu_millicharge}
\end{figure}

Dark matter direct detection experiments are sensitive to all of the neutrino millicharges. Note that due to the choice $\sin^2\theta_{23} = 0.5$ the bounds on the muon and tau neutrino millicharges are going to be the same for the same reason as explained in the previous section. The results from our analyses are presented in Figs.~\ref{fig:nu_millicharge_2D} and~\ref{fig:nu_millicharge} and in Table~\ref{tab:nu_millicharge}. 
An interesting feature can be seen in Fig.~\ref{fig:nu_millicharge_2D}. While the preferred regions of the current experiments have circular shape (left panel), the region of DARWIN does not (right panel). The reason for these shapes is the following: If one substitutes Eq.~\eqref{eq:gV-EC} into Eq.~\eqref{eq:ES-SM}, one sees that the cross section depends on terms which are proportional to $q_{\nu_\alpha}$ and to $q_{\nu_\alpha}^2$. In the case of $q_{\nu_{\mu/\tau}}$, the dominating new physics contribution to the cross section comes always from the terms which are proportional to $q_{\nu_{\mu/\tau}}^2$. Also in the case of electron neutrinos the dominating contribution comes from the terms which are proportional to $q_{\nu_e}^2$ when considering current experiments. Therefore, no cancellations can occur among the new physics parameters and the preferred regions obtain the circular shape. However, in the case of DARWIN, which is sensitive to smaller millicharges, the contributions of both terms are of similar strength for electron neutrinos. Hence, correlations can appear among the parameters, which is reflected in the shape of the allowed regions. In our analyses we have taken the possible correlations among the parameters into account.

As can be seen in Figs.~\ref{fig:nu_millicharge_2D} and~\ref{fig:nu_millicharge}, again the PandaX-4T bound is weaker than the one from LZ, while this one is slightly weaker than that from XENONnT. Due to the conservative approach of background treatment, our bounds are again slightly weaker than those obtained in Refs.~\cite{AtzoriCorona:2022jeb,A:2022acy}. 

At 90\% confidence level we obtain from the combined analysis: 
\begin{eqnarray}
    q_{\nu_{e}} &\in& (-2.0,7.0)\times10^{-13}~e\,, \\
    q_{\nu_{\mu/\tau}} &\in& (-7.5,7.3)\times10^{-13}~e\,, \\
    |q_{\nu_{e\mu/e\tau}}| &<& 4.1\times10^{-13}~e\,, \\
    |q_{\nu_{\mu\tau}}| &<& 5.2\times10^{-13}~e\,.
\end{eqnarray}
Our bound on $q_{\nu_{e}}$ is stronger than the bounds from the analyses of the data from TEXONO
($|q_{\nu_{e}}| < 1.0 \times 10^{-12} \, e$ at 90\% C.L.)~\cite{TEXONO:2002pra,TEXONO:2006xds,Gninenko:2006fi,Chen:2014dsa} or GEMMA
($|q_{\nu_{e}}| < 1.5 \times 10^{-12} \, e$ at 90\% C.L.)~\cite{Studenikin:2013my,Beda:2012zz}.
Regarding $q_{\nu_{\mu/\tau}}$, the bound is more than one order of magnitude more stringent than that obtained from solar neutrinos by the XMASS collaboration ($|q_{\nu_{\mu/\tau}}| < 1.1 \times 10^{-11} \, e$ at 90\% C.L.)~\cite{XMASS:2020zke}.
In the case of the non-diagonal millicharges, we improve the DRESDEN-II \cevns bound~\cite{AtzoriCorona:2022qrf} on $|q_{\nu_{e\mu/e\tau}}|$ by more than one order of magnitude and
we improve the \cevns COHERENT bound~\cite{AtzoriCorona:2022qrf} on $|q_{\nu_{\mu\tau}}|$
by about three orders of magnitude.

All of these bounds will be further improved significantly by the DARWIN experiment, up to a factor of more than 20 in the most optimistic scenario in the case of $q_{\nu_{e}}$, making dark matter direct detection experiments again competitive with astrophysical observations
which constrain the neutrino millicharges below a few $ 10^{-14} \, e $~\cite{Raffelt:1999gv}.

\begin{table}[t]
\centering
\begin{tabular}{|c||c|c|c|c|}
\hline
Experiment & $q_{\nu_{e}} ~[10^{-13}~e]$  & $q_{\nu_{\mu}} ~[10^{-13}~e]$ & $|q_{\nu_{e\mu/e\tau}}| ~[10^{-13}~e]$ & 
$|q_{\nu_{\mu\tau}}| ~[10^{-13}~e]$\\
\hline
PandaX-4T & $(-12.6,16.4)$ & $(-22.3,22.2)$ & $<12.2$& $<15.7$\\
\hline
LZ & $(-4.6,9.9)$ & $(-11.5,11.3)$ & $<6.3$  & $<8.1$  \\
\hline
XENONnT & $(-2.5,7.4)$ & $(-8.1,8.0)$ & $<4.4$  & $<5.7$  \\
\hline
combined & $(-2.0,7.0)$ & $(-7.5,7.3)$ & $<4.1$  & $<5.2$  \\
\hline
\hline
DARWIN 30~ty & $(-0.4,1.0)$ & $(-4.1,4.1)$ & $<2.3$  & $<2.9$  \\
\hline
DARWIN 300~ty & $(-0.2,0.4)$ & $(-2.4,2.5)$ & $<1.3$  & $<1.7$  \\
\hline
DARWIN 300~ty depl. & $(-0.1,0.3)$ & $(-2.2,2.3)$ & $<1.2$  & $<1.6$  \\
\hline
\end{tabular}
\caption{The 90\% bounds ($\Delta\chi^2=2.71$) that can be obtained on the different neutrino millicharges.}
\label{tab:nu_millicharge}
\end{table}

\subsection{Neutrino charge radius}
\label{sec:nu_chargeradius}

\begin{figure}
\begin{subfigure}{0.48\textwidth}
    \includegraphics[width=\textwidth]{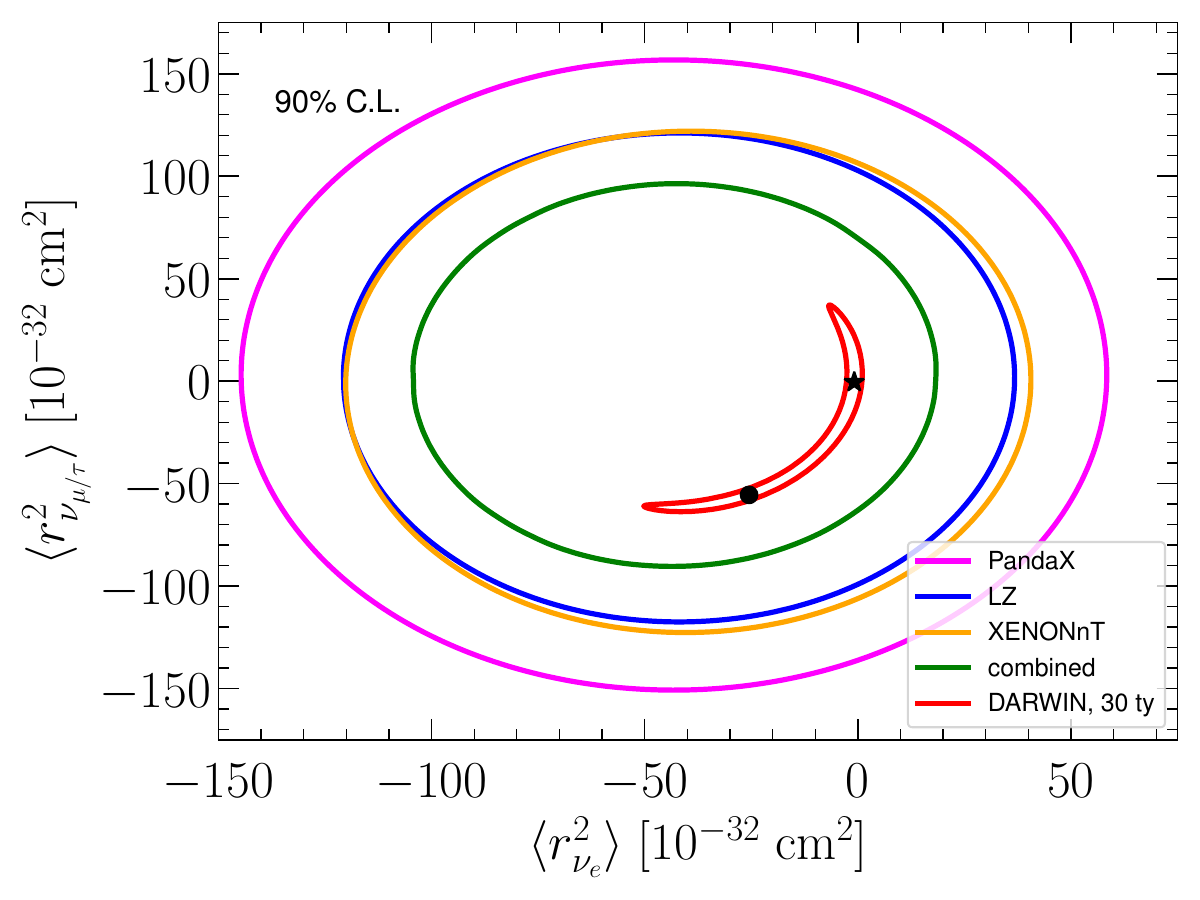}
\end{subfigure}
\hfill
\begin{subfigure}{0.48\textwidth}
    \includegraphics[width=\textwidth]{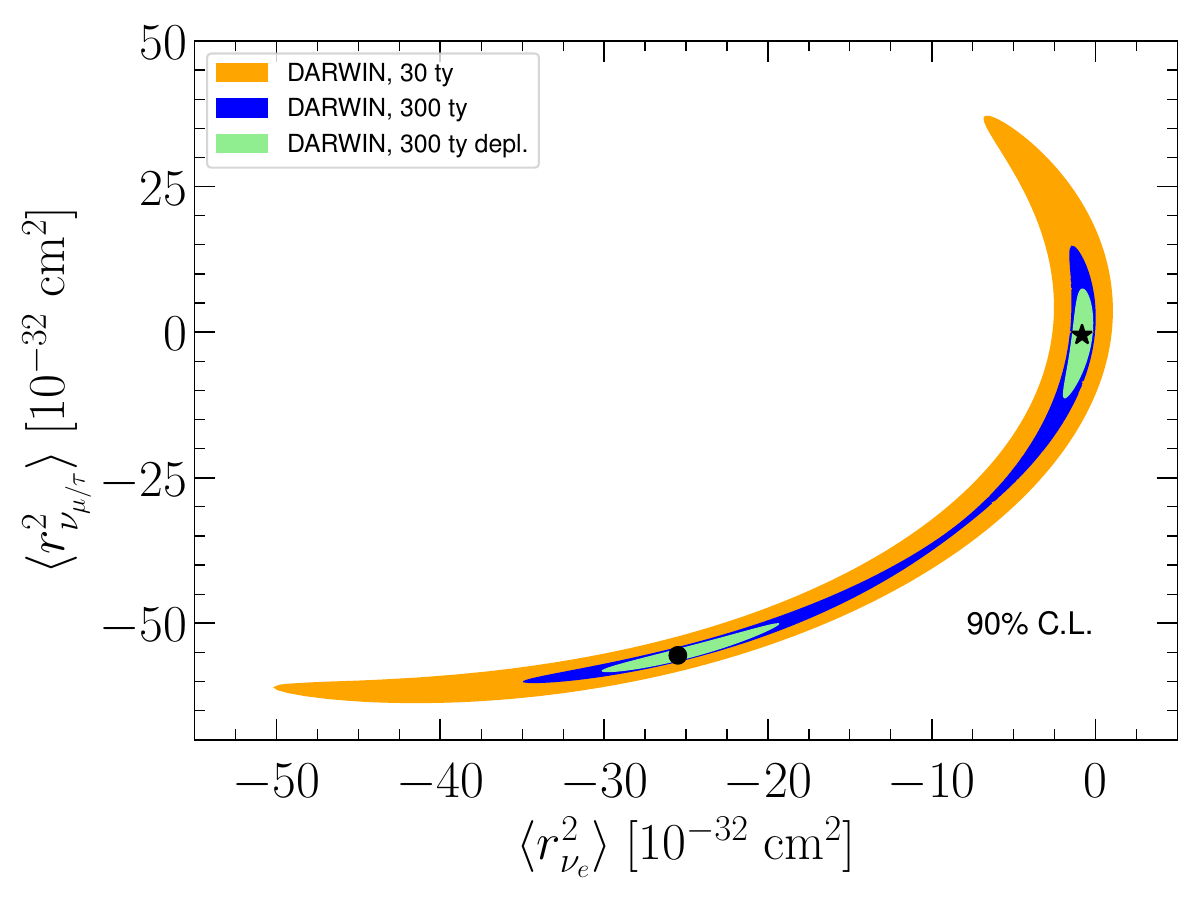}
    %caption
    %label
\end{subfigure}
\hfill
\caption{Left: The 90\% C.L. (2 d.o.f.) allowed regions in the $\langle r_{\nu_{e}}^2\rangle-\langle r_{\nu_{\mu}}^2\rangle$-plane from PandaX (magenta), LZ (blue), XENONnT (orange) and from the combination of all experiments (green) in comparison with the DARWIN sensitivity assuming 30~ty of exposure. Right: The 90\% C.L. (2 d.o.f.) expected sensitivity of DARWIN in the $\langle r_{\nu_{e}}^2\rangle-\langle r_{\nu_{\mu}}^2\rangle$-plane for different exposures and background assumptions. The star denotes the SM value, while the black circle denotes the position of a secondary minimum found in the analysis.}
\label{fig:nu_chargeradius_2D}
\end{figure}

Finally, we can use the data to place bounds on the neutrino charge radii. Regarding the charge radii, similar correlations as discussed in the context of millicharges have to be taken into account, as can be seen in Fig.~\ref{fig:nu_chargeradius_2D}. 

\begin{figure}
\begin{subfigure}{0.48\textwidth}
    \includegraphics[width=\textwidth]{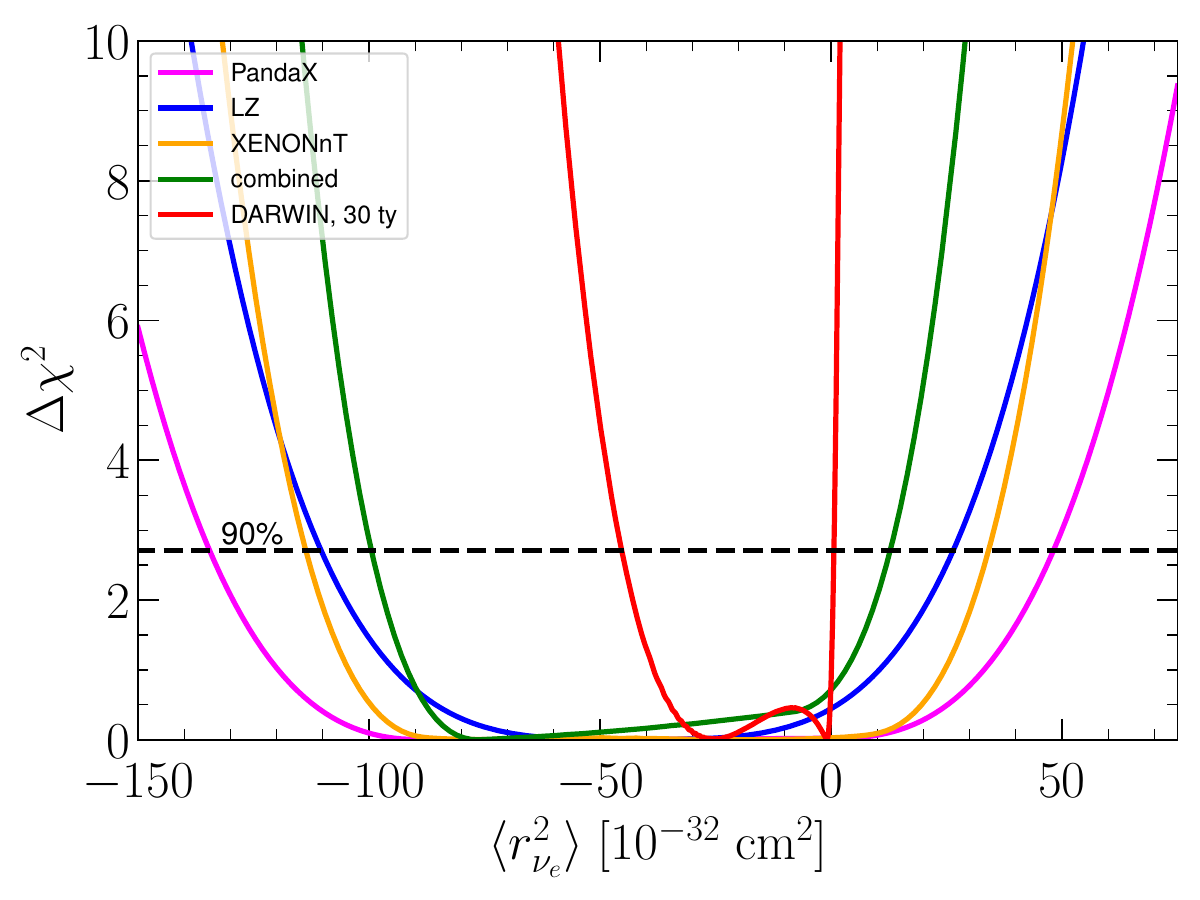}
    %caption
    %label
\end{subfigure}
\hfill
\hfill
\begin{subfigure}{0.48\textwidth}
    \includegraphics[width=\textwidth]{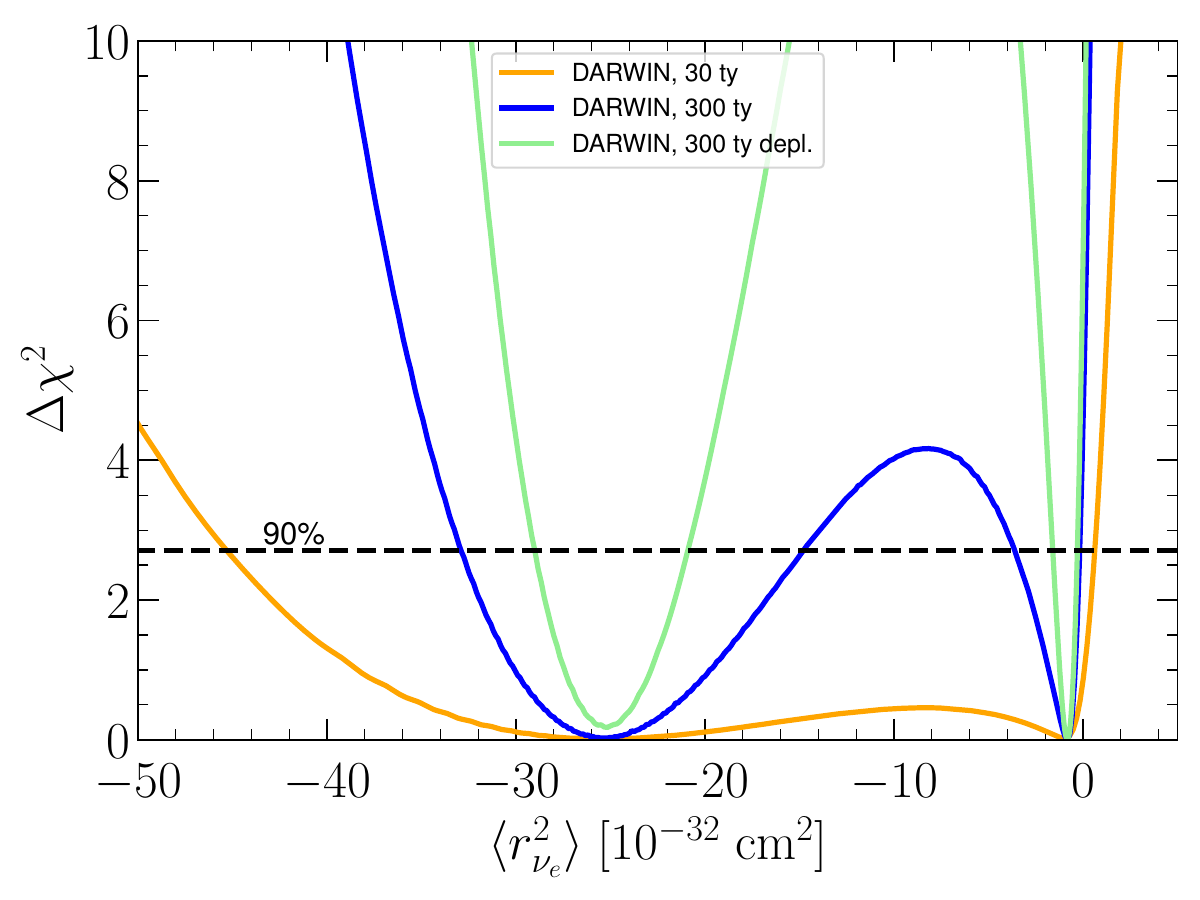}
    %caption
    %label
\end{subfigure}
\hfill
\hfill
\begin{subfigure}{0.48\textwidth}
    \includegraphics[width=\textwidth]{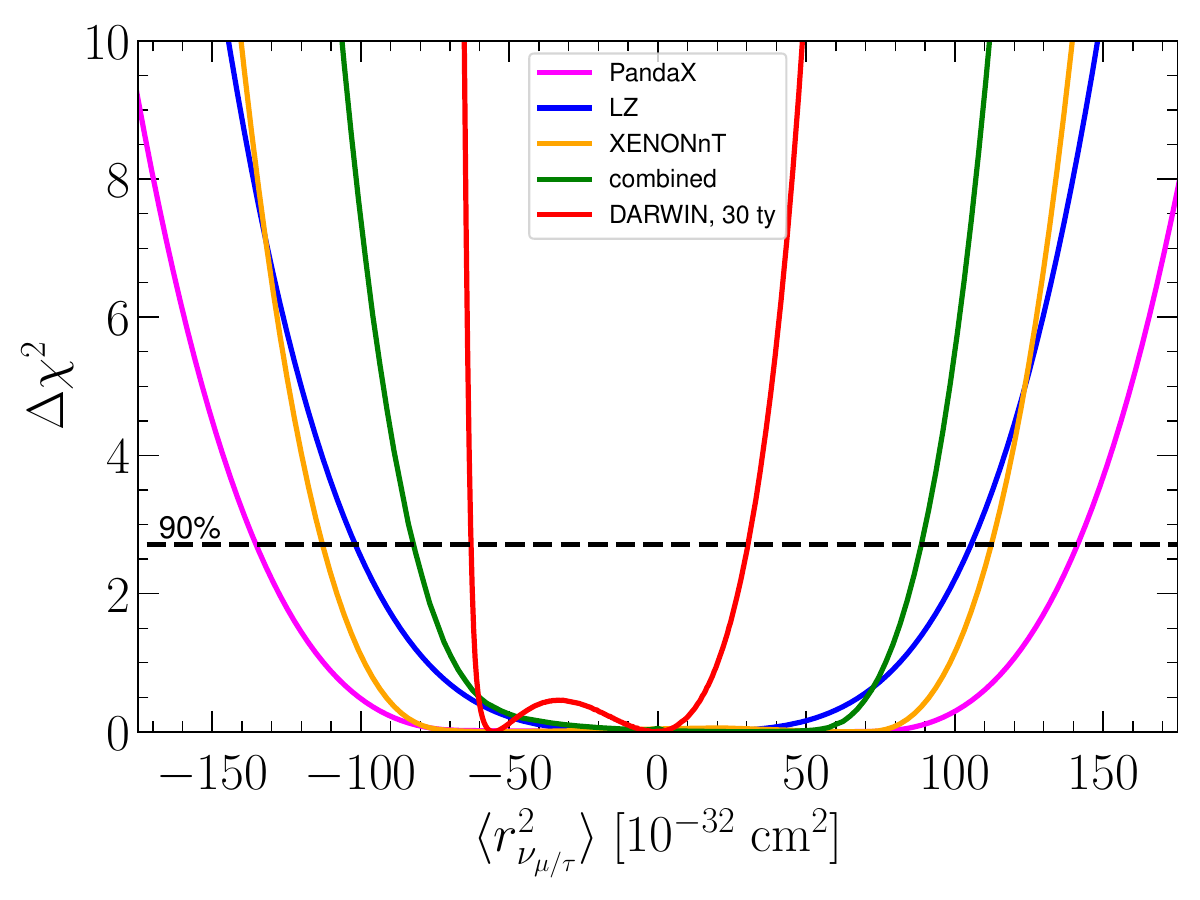}
    %caption
    %label
\end{subfigure}
\hfill
\hfill
\begin{subfigure}{0.48\textwidth}
    \includegraphics[width=\textwidth]{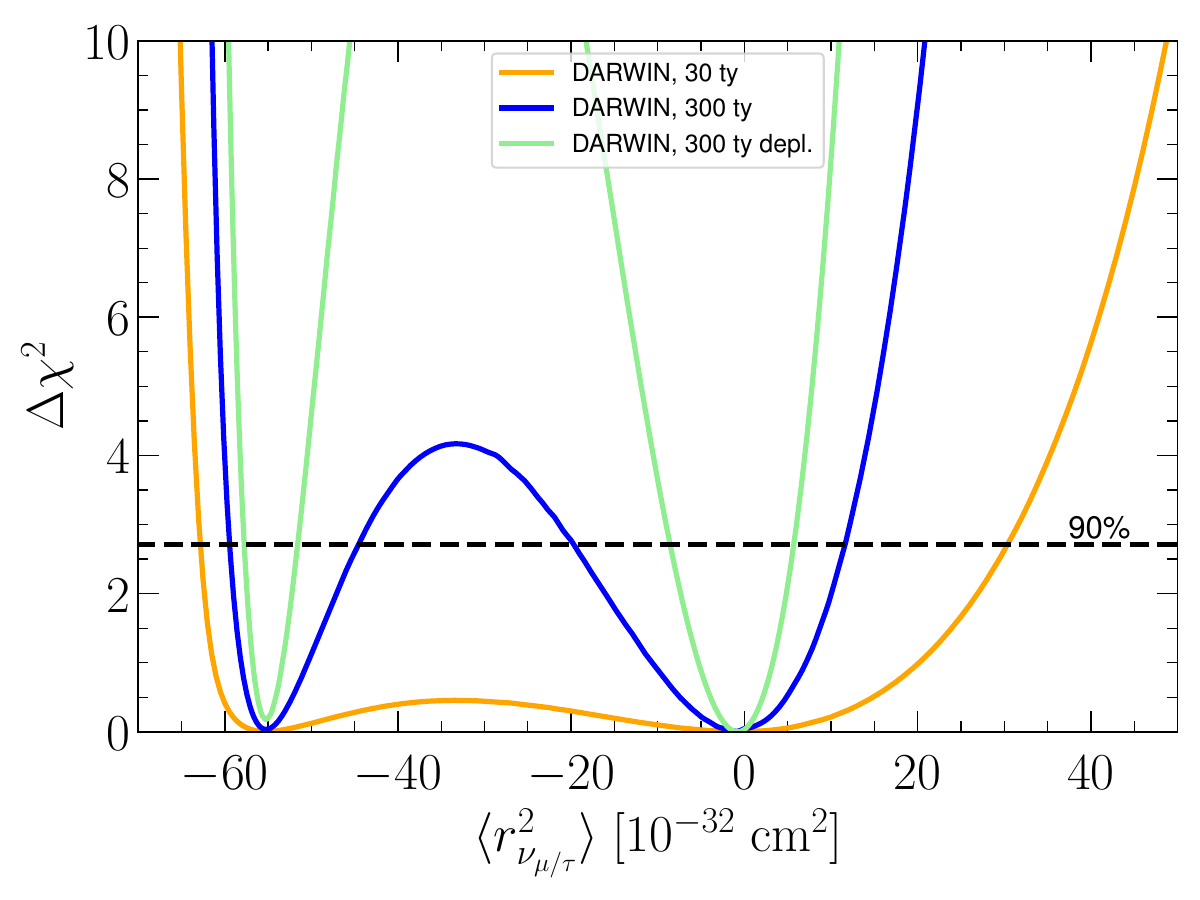}
    %caption
    %label
\end{subfigure}
\hfill
\caption{$\Delta\chi^2$ profiles for the diagonal charge radii obtained from the analyses of current data (left panels) and the DARWIN sensitivity (right panels). For comparison, we also show in the left panels the DARWIN-sensitivity corresponding to an exposure of 30~ty.}
\label{fig:nu_chargeradius_1D_diag}
\end{figure}

\begin{figure}
\begin{subfigure}{0.48\textwidth}
    \includegraphics[width=\textwidth]{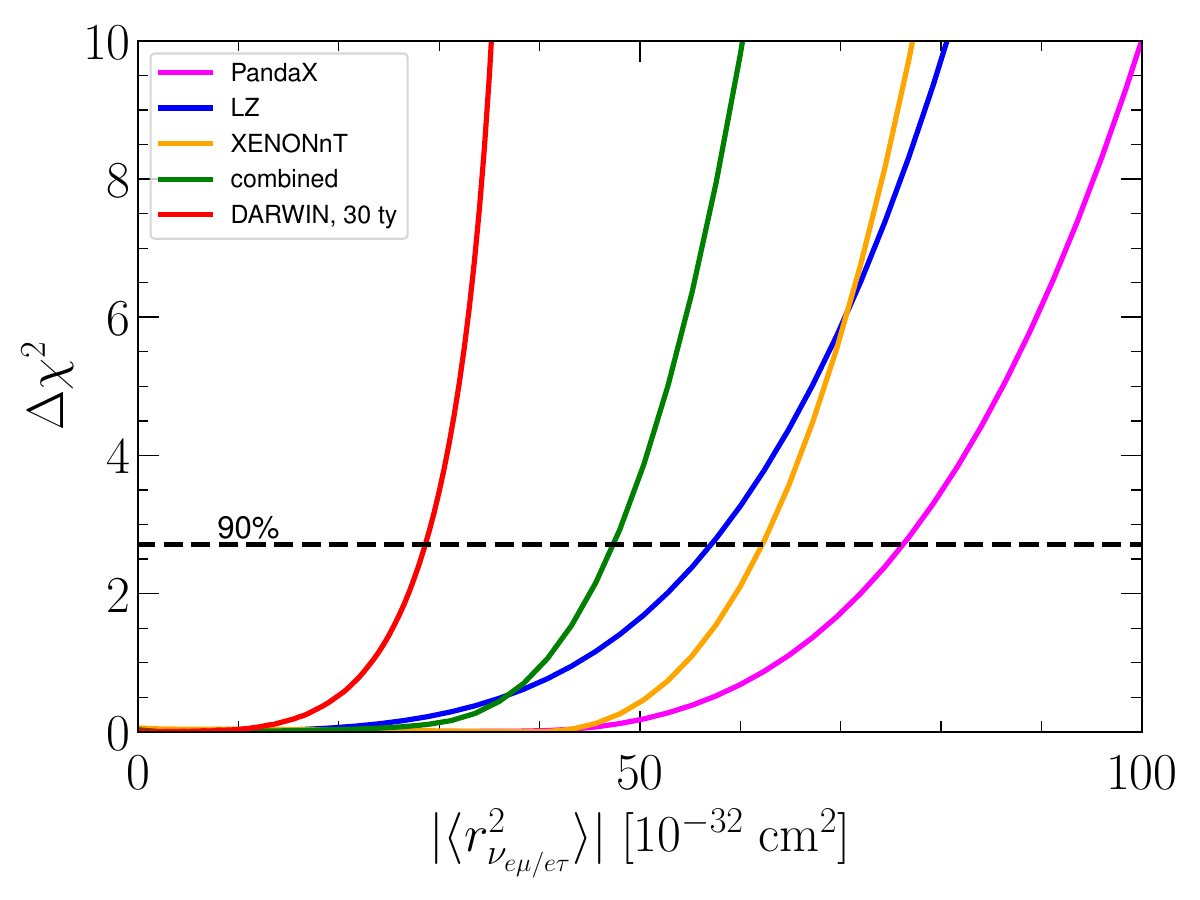}
    %caption
    %label
\end{subfigure}
\hfill
\hfill
\begin{subfigure}{0.48\textwidth}
    \includegraphics[width=\textwidth]{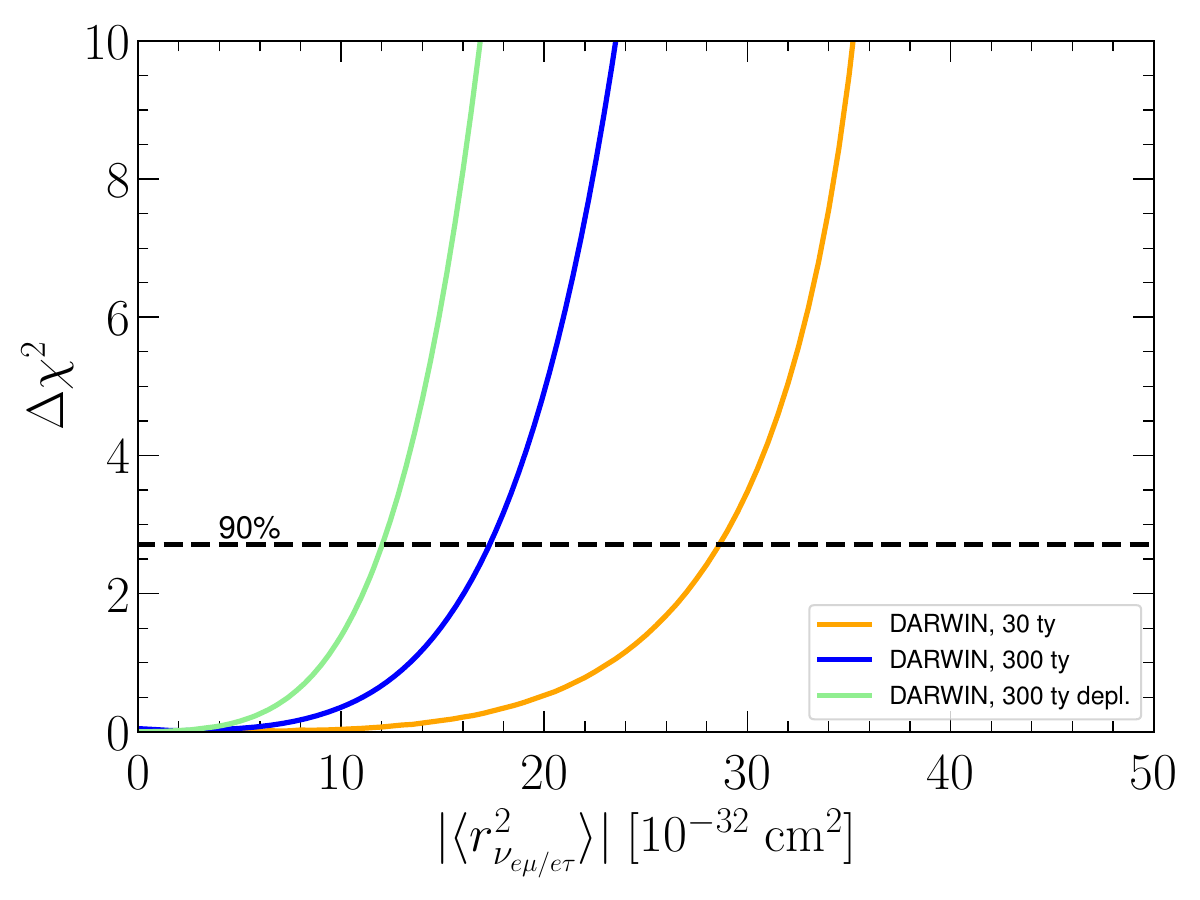}
    %caption
    %label
\end{subfigure}
\hfill
\hfill
\begin{subfigure}{0.48\textwidth}
    \includegraphics[width=\textwidth]{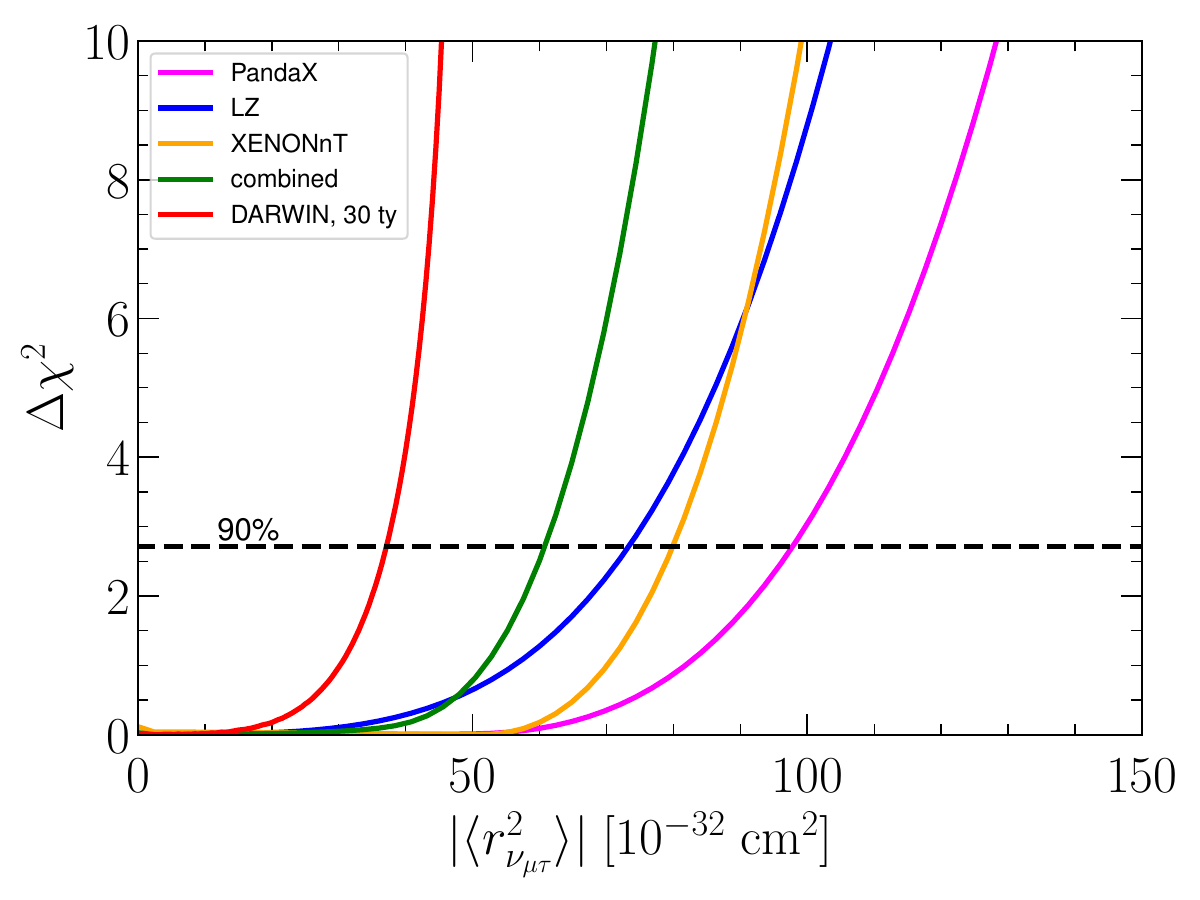}
    %caption
    %label
\end{subfigure}
\hfill
\hfill
\begin{subfigure}{0.48\textwidth}
    \includegraphics[width=\textwidth]{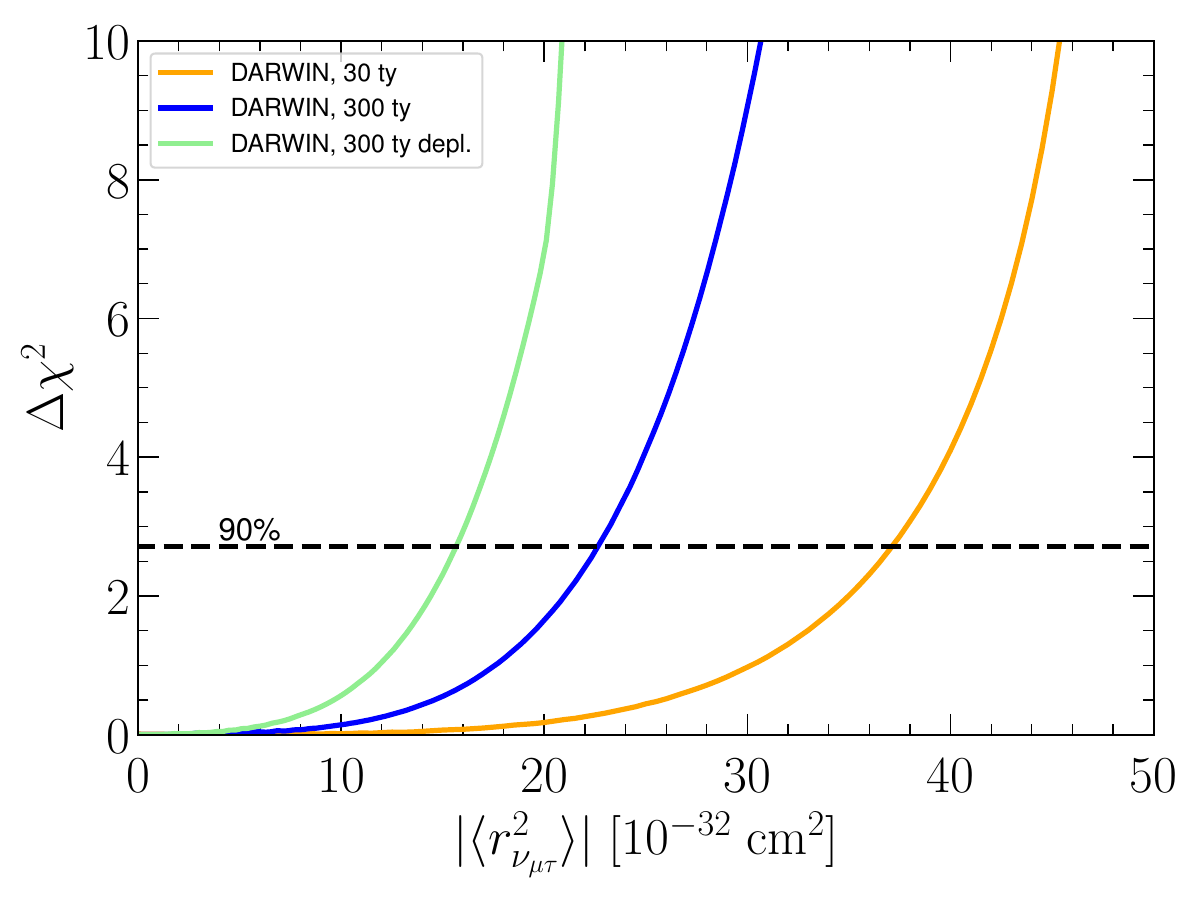}
    %caption
    %label
\end{subfigure}
\hfill
\caption{$\Delta\chi^2$ profiles for the non-diagonal charge radii obtained from the analyses of current data (left panels) and the DARWIN sensitivity (right panels). For comparison, we also show in the left panels the DARWIN-sensitivity corresponding to an exposure of 30~ty.}
\label{fig:nu_chargeradius_1D_nondiag}
\end{figure}

The results of the analyses of PandaX-4T, LZ and XENONnT and from the combined analysis are shown in Table~\ref{tab:nu_chargeradii}, in the left panels of Figs.~\ref{fig:nu_chargeradius_2D} and~\ref{fig:nu_chargeradius_1D_diag} for the diagonal charge radii, and Fig.~\ref{fig:nu_chargeradius_1D_nondiag} for the non-diagonal ones.
As can be seen in Table~\ref{tab:nu_chargeradii}, the bound on $\langle r_{\nu_{e}}^2\rangle$ is stronger than that on $\langle r_{\nu_{\mu/\tau}}^2\rangle$. Unfortunately, current dark matter direct detection experiments are not competitive with the bounds from other experiments~\cite{AtzoriCorona:2022qrf,TEXONO:2009knm,Ahrens:1990fp} for neither of the two charge radii
and they are far from the Standard Model values in Eqs.~\eqref{reSM}--\eqref{rtSM}.
Also the bounds on the non-diagonal parameters are weaker than those obtained in CE$\nu$NS experiments~\cite{AtzoriCorona:2022qrf}.

However, we have found that DARWIN will improve significantly the allowed region of parameter space. In the left panel of Fig.~\ref{fig:nu_chargeradius_2D} we show the expected allowed region from DARWIN for a 30~ty exposure. 
As can be seen, DARWIN could significantly reduce the volume of the allowed parameter space. With this relatively small exposure it would be possible to set the strongest upper limit on $\langle r_{\nu_{e}}^2\rangle$, although the lower limit would remain weaker than the most stringent current bound obtained in the TEXONO experiment
($\langle r_{\nu_{e}}^2\rangle \in (-4.2,6.6) \times 10^{-32}~\text{cm}^2$ at 90\% C.L.)~\cite{TEXONO:2009knm}. The size of the sensitivity region shrinks when considering a larger exposure or the better background model, as shown in the right panel of Fig.~\ref{fig:nu_chargeradius_2D}. 
In the analysis with a 300~ty exposure, values around $\langle r_{\nu_{e}}^2\rangle\approx -10\times 10^{-32}~\text{cm}^2$ and $\langle r_{\nu_{\mu/\tau}}^2\rangle\approx -35\times 10^{-32}~\text{cm}^2$ become more disfavored, but there is still a secondary minimum present, as can be seen in the right panels of Figs.~\ref{fig:nu_chargeradius_2D} and~\ref{fig:nu_chargeradius_1D_diag}. 
The secondary solution remains even when considering the better background model. The bounds that could be obtained at 90\% confidence level are
\begin{eqnarray}
    \langle r_{\nu_{e}}^2\rangle &\in& (-45.3,0.6)\times 10^{-32}~\text{cm}^2, \text{ DARWIN 30~ty}\,, \\
    \langle r_{\nu_{e}}^2\rangle &\in& \left\{(-32.9,-14.8)~\&~(-3.6,-0.2)\right\}\times 10^{-32}~\text{cm}^2, \text{ DARWIN 300~ty}\,, \\
    \langle r_{\nu_{e}}^2\rangle &\in& \left\{(-29.1,-20.7)~\&~(-1.6,-0.3)\right\}\times 10^{-32}~\text{cm}^2, \text{ DARWIN 300~ty, depleted}\,. 
\end{eqnarray}
Thus, DARWIN 300~ty could improve the current best limit on $\langle r_{\nu_{e}}^2\rangle$
obtained in the TEXONO experiment
($\langle r^2 \rangle_{\nu_{e}}\in(-4.2,6.6)\times10^{-32}$ at 90\% C.L.)~\cite{TEXONO:2009knm}
(taking into account that the secondary solution is excluded by TEXONO and other bounds; see, e.g., Refs.~\cite{Giunti:2014ixa,Giunti:2015gga,Giunti:2022aea,ParticleDataGroup:2022pth})
and could indicate that $\langle r_{\nu_{e}}^2\rangle$ is negative. 
This would be the first indication of a non-zero value of a neutrino charge radius,
in agreement with the Standard Model prediction~\eqref{reSM}.

In the lower panels of Fig.~\ref{fig:nu_chargeradius_1D_diag} we show the sensitivity of DARWIN to the charge radius $\langle r_{\nu_{\mu}}^2\rangle$. Although not as strong as for $\langle r_{\nu_{e}}^2\rangle$, DARWIN could provide strong bounds on this quantity, which read at 90\% confidence level
\begin{eqnarray}
    \langle r_{\nu_{\mu}}^2\rangle &\in& (-62.9,30.4)\times 10^{-32}~\text{cm}^2, \text{ DARWIN 30~ty}\,, \\
    \langle r_{\nu_{\mu}}^2\rangle &\in& \left\{(-59.5,-44.6)~\&~(-19.9,11.7)\right\}\times 10^{-32}~\text{cm}^2, \text{ DARWIN 300~ty}\,, \\
    \langle r_{\nu_{\mu}}^2\rangle &\in& \left\{(-57.8,-51.4)~\&~(-8.6,5.7)\right\}\times 10^{-32}~\text{cm}^2, \text{ DARWIN 300~ty, depleted}\,.
\end{eqnarray}
Such bounds would be complementary to the bounds obtained in other experiments~\cite{AtzoriCorona:2022qrf,Ahrens:1990fp,CHARM-II:1994aeb,Khan:2017djo}. 
Note that the secondary minimum obtained in our DARWIN analyses requires both charge radii to be different from the Standard Model values.
We could eliminate the secondary minimum by using external constraints on 
$\langle r_{\nu_{e}}^2\rangle$
(e.g the TEXONO bound~\cite{TEXONO:2009knm})
or
$\langle r_{\nu_{\mu}}^2\rangle$ (e.g. the one from Ref.~\cite{Ahrens:1990fp}).

Regarding the transition charge radii, the DARWIN sensitivity is shown in comparison with the current bounds in Fig.~\ref{fig:nu_chargeradius_1D_nondiag} and in the last two columns of Table~\ref{tab:nu_chargeradii}. In this case the bounds that DARWIN could set assuming an exposure of 30~ty are similar in strength to those obtained with \cevns in Ref.~\cite{AtzoriCorona:2022qrf}. Only with a larger exposure and a better background model the current bounds could be improved by a factor of about 2--3.

%Note that as in the previous cases a very similar bound\footnote{In this case the bound is not exactly the same due to the different Standard Model values of $\langle r_{\nu_{\mu}}^2\rangle$ and $\langle r_{\nu_{\tau}}^2\rangle$. However, this difference is so small in comparison with the sensitivity of the experiment that for all practical purpose one can argue that the bound is the same.} can be set for $\langle r_{\nu_{\tau}}^2\rangle$. In this case DARWIN could improve the bound from low-energy solar neutrinos found in Ref.~\cite{Khan:2017djo}. 

\begin{table}[t]
\centering
\begin{tabular}{|c||c|c|c|c|}
\hline
Experiment & $\langle r_{\nu_{e}}^2\rangle ~[10^{-32}]$~cm$^2$  & $\langle r_{\nu_{\mu/\tau}}^2\rangle ~[10^{-32}]$~cm$^2$ & $|\langle r_{\nu_{e\mu/e\tau}}^2\rangle| ~[10^{-32}]$~cm$^2$ & 
$|\langle r_{\nu_{\mu\tau}}^2\rangle| ~[10^{-32}]$~cm$^2$\\
\hline
PandaX-4T & $(-134.5,48.2)$ & $(-135.3,141.3)$ & $<76.2$& $<97.8$\\
\hline
LZ & $(-110.4,26.4)$ & $(-101.8,105.5)$ & $<57.1$   & $<73.3$  \\
\hline
XENONnT & $(-113.7,34.1)$ & $(-112.9,112.3)$ & $<62.2$  & $<79.9$  \\
\hline
combined & $(-99.5,12.8)$ & $(-82.2,88.7)$ & $<47.3$  & $<60.7$  \\
\hline
\hline
DARWIN 30~ty & $(-45.3,0.6)$ & $(-62.9,30.4)$ & $<28.6$  & $<37.9$  \\
\hline
DARWIN 300~ty & \makecell{$(-32.9,-14.8)$ \\ \& $(-3.6,-0.2)$} & \makecell{$(-59.5,-44.6)$ \\ \& $(-19.9,11.7)$} & $<28.6$  & $<37.9$  \\
\hline
DARWIN 300~ty depl. & \makecell{ $(-29.1,-20.7)$ \\ \&$(-1.6,-0.3)$} & \makecell{  $(-57.8,-51.4)$\&$(-8.6,5.7)$} & $<12.0$  & $<15.7$  \\
\hline
\end{tabular}
\caption{The 90\% bounds ($\Delta\chi^2=2.71$) that can be obtained on the different neutrino charge radii.}
\label{tab:nu_chargeradii}
\end{table}

\section{Conclusions}
\label{sec:conclusions}

In this paper we have analyzed the electron recoil data from the dark matter direct detection experiments PandaX-4T, LZ and XENONnT to place constraints on neutrino electromagnetic properties. We also explored the sensitivity of the future experiment DARWIN. We implemented a more realistic treatment of systematic uncertainties than in other phenomenological analyses previously performed in the literature, following closer what is done by the collaborations. 

We have set strong limits on the neutrino magnetic moments and on the neutrino millicharges. These are still weaker than those from astrophysical probes. However, we have shown that the next generation experiment DARWIN will put the sensitivity of dark matter direct detection experiments into the same ballpark. 

We have shown that in the case of the neutrino charge radii current dark matter direct detection experiments are not competitive with other types of experiments. However, this situation will change dramatically for DARWIN. Under ideal circumstances a measurement of $\langle r_{\nu_{e}}^2\rangle$ could be possible for the first time.

An interesting feature of dark matter direct detection experiments which can detect solar neutrinos is that part of the neutrinos arrives as $\nu_\tau$, hence allowing to set bounds also on the $\nu_\tau$ related quantities, which is not possible with other experiments.

Summarizing, we have shown that current and particularly future dark matter direct detection experiments provide a powerful tool to test neutrino electromagnetic properties.

\section*{Acknowledgments}
We are very thankful to Dimitris Papoulias for helful discussions. C.G. and C.A.T. acknowledge support from {\sl Departments of Excellence} grant awarded by MIUR and the research grant {\sl TAsP (Theoretical Astroparticle Physics)} funded by Istituto Nazionale di Fisica Nucleare (INFN).

%\bibliographystyle{utphys}
%\bibliography{bibliography}  

\begin{thebibliography}{10}

\bibitem{Giunti:2014ixa}
C.~Giunti and A.~Studenikin, ``{Neutrino electromagnetic interactions: a window
  to new physics},'' \href{http://dx.doi.org/10.1103/RevModPhys.87.531}{{\em
  Rev. Mod. Phys.} {\bfseries 87} (2015) 531},
  \href{http://arxiv.org/abs/1403.6344}{{\ttfamily arXiv:1403.6344 [hep-ph]}}.

\bibitem{Giunti:2015gga}
C.~Giunti, K.~A. Kouzakov, Y.-F. Li, A.~V. Lokhov, A.~I. Studenikin, and
  S.~Zhou, ``{Electromagnetic neutrinos in laboratory experiments and
  astrophysics},'' \href{http://dx.doi.org/10.1002/andp.201500211}{{\em Annalen
  Phys.} {\bfseries 528} (2016) 198--215},
  \href{http://arxiv.org/abs/1506.05387}{{\ttfamily arXiv:1506.05387
  [hep-ph]}}.

\bibitem{PandaX:2022ood}
{\bfseries PandaX} Collaboration, D.~Zhang {\em et~al.}, ``{Search for Light
  Fermionic Dark Matter Absorption on Electrons in PandaX-4T},''
  \href{http://dx.doi.org/10.1103/PhysRevLett.129.161804}{{\em Phys. Rev.
  Lett.} {\bfseries 129} no.~16, (2022) 161804},
  \href{http://arxiv.org/abs/2206.02339}{{\ttfamily arXiv:2206.02339
  [hep-ex]}}.

\bibitem{LZ:2022lsv}
{\bfseries LZ} Collaboration, J.~Aalbers {\em et~al.}, ``{First Dark Matter
  Search Results from the LUX-ZEPLIN (LZ) Experiment},''
  \href{http://dx.doi.org/10.1103/PhysRevLett.131.041002}{{\em Phys. Rev.
  Lett.} {\bfseries 131} no.~4, (2023) 041002},
  \href{http://arxiv.org/abs/2207.03764}{{\ttfamily arXiv:2207.03764
  [hep-ex]}}.

\bibitem{XENON:2022ltv}
{\bfseries XENON} Collaboration, E.~Aprile {\em et~al.}, ``{Search for New
  Physics in Electronic Recoil Data from XENONnT},''
  \href{http://dx.doi.org/10.1103/PhysRevLett.129.161805}{{\em Phys. Rev.
  Lett.} {\bfseries 129} no.~16, (2022) 161805},
  \href{http://arxiv.org/abs/2207.11330}{{\ttfamily arXiv:2207.11330
  [hep-ex]}}.

\bibitem{DARWIN:2020bnc}
{\bfseries DARWIN} Collaboration, J.~Aalbers {\em et~al.}, ``{Solar neutrino
  detection sensitivity in DARWIN via electron scattering},''
  \href{http://dx.doi.org/10.1140/epjc/s10052-020-08602-7}{{\em Eur. Phys. J.
  C} {\bfseries 80} no.~12, (2020) 1133},
  \href{http://arxiv.org/abs/2006.03114}{{\ttfamily arXiv:2006.03114
  [physics.ins-det]}}.

\bibitem{Giunti:2022aea}
C.~Giunti, J.~Gruszko, B.~Jones, L.~Kaufman, D.~Parno, and A.~Pocar, ``{Report
  of the Topical Group on Neutrino Properties for Snowmass 2021},''
  \href{http://arxiv.org/abs/arXiv:2209.03340}{{\ttfamily arXiv:2209.03340
  [hep-ph]}}.

\bibitem{bahcall_web}
J.~Bahcall. \url{http://www.sns.ias.edu/~jnb/SNdata/sndata.html}.

\bibitem{Bahcall:1987jc}
J.~N. Bahcall and R.~K. Ulrich, ``{Solar Models, Neutrino Experiments and
  Helioseismology},'' \href{http://dx.doi.org/10.1103/RevModPhys.60.297}{{\em
  Rev. Mod. Phys.} {\bfseries 60} (1988) 297--372}.

\bibitem{Bahcall:1994cf}
J.~N. Bahcall, ``{The Be-7 solar neutrino line: A Reflection of the central
  temperature distribution of the sun},''
  \href{http://dx.doi.org/10.1103/PhysRevD.49.3923}{{\em Phys. Rev. D}
  {\bfseries 49} (1994) 3923--3945},
  \href{http://arxiv.org/abs/astro-ph/9401024}{{\ttfamily
  arXiv:astro-ph/9401024}}.

\bibitem{Bahcall:1996qv}
J.~N. Bahcall, E.~Lisi, D.~E. Alburger, L.~De~Braeckeleer, S.~J. Freedman, and
  J.~Napolitano, ``{Standard neutrino spectrum from B-8 decay},''
  \href{http://dx.doi.org/10.1103/PhysRevC.54.411}{{\em Phys. Rev. C}
  {\bfseries 54} (1996) 411--422},
  \href{http://arxiv.org/abs/nucl-th/9601044}{{\ttfamily
  arXiv:nucl-th/9601044}}.

\bibitem{Villante:2020adi}
F.~L. Villante and A.~Serenelli, ``{The relevance of nuclear reactions for
  Standard Solar Models construction},''
  \href{http://dx.doi.org/10.3389/fspas.2020.618356}{{\em Front. Astron. Space
  Sci.} {\bfseries 7} (2021) 112},
  \href{http://arxiv.org/abs/2101.03077}{{\ttfamily arXiv:2101.03077
  [astro-ph.SR]}}.

\bibitem{deSalas:2020pgw}
P.~F. de~Salas, D.~V. Forero, S.~Gariazzo, P.~Martinez-Mirave, O.~Mena, C.~A.
  Ternes, M.~Tortola, and J.~W.~F. Valle, ``{2020 Global reassessment of the
  neutrino oscillation picture},'' {\em JHEP} {\bfseries 2021} (2020) 071,
  \href{http://arxiv.org/abs/arXiv:2006.11237}{{\ttfamily arXiv:2006.11237
  [hep-ph]}}.

\bibitem{Esteban:2020cvm}
I.~Esteban, M.~C. Gonzalez-Garcia, M.~Maltoni, T.~Schwetz, and A.~Zhou, ``{The
  fate of hints: updated global analysis of three-flavor neutrino
  oscillations},'' \href{http://dx.doi.org/10.1007/JHEP09(2020)178}{{\em JHEP}
  {\bfseries 09} (2020) 178},
  \href{http://arxiv.org/abs/arXiv:2007.14792}{{\ttfamily arXiv:2007.14792
  [hep-ph]}}.

\bibitem{Capozzi:2021fjo}
F.~Capozzi, E.~{Di Valentino}, E.~Lisi, A.~Marrone, A.~Melchiorri, and
  A.~Palazzo, ``{The unfinished fabric of the three neutrino paradigm},'' {\em
  Phys.Rev.D} {\bfseries 104} (7, 2021) 083031,
  \href{http://arxiv.org/abs/arXiv:2107.00532}{{\ttfamily arXiv:2107.00532
  [hep-ph]}}.

\bibitem{ParticleDataGroup:2022pth}
{\bfseries Particle Data Group} Collaboration, R.~L. Workman, ``{Review of
  Particle Physics},'' {\em PTEP} {\bfseries 2022} (2022) 083C01.

\bibitem{Fayans:2000ns}
S.~A. Fayans, L.~A. Mikaelyan, and V.~V. Sinev, ``{Weak and magnetic inelastic
  scattering of anti-neutrinos on atomic electrons},''
  \href{http://dx.doi.org/10.1134/1.1398940}{{\em Phys. Atom. Nucl.} {\bfseries
  64} (2001) 1475--1480}, \href{http://arxiv.org/abs/hep-ph/0004158}{{\ttfamily
  hep-ph/0004158}}.

\bibitem{AtzoriCorona:2022jeb}
M.~Atzori~Corona, W.~M. Bonivento, M.~Cadeddu, N.~Cargioli, and F.~Dordei,
  ``{New constraint on neutrino magnetic moment and neutrino millicharge from
  LUX-ZEPLIN dark matter search results},''
  \href{http://dx.doi.org/10.1103/PhysRevD.107.053001}{{\em Phys. Rev. D}
  {\bfseries 107} no.~5, (2023) 053001},
  \href{http://arxiv.org/abs/2207.05036}{{\ttfamily arXiv:2207.05036
  [hep-ph]}}.

\bibitem{Beda:2012zz}
A.~G. Beda, V.~B. Brudanin, V.~G. Egorov, D.~V. Medvedev, V.~S. Pogosov, M.~V.
  Shirchenko, and A.~S. Starostin, ``{The results of search for the neutrino
  magnetic moment in GEMMA experiment},''
  \href{http://dx.doi.org/10.1155/2012/350150}{{\em Adv. High Energy Phys.}
  {\bfseries 2012} (2012) 350150}.

\bibitem{Fujikawa:1980yx}
K.~Fujikawa and R.~Shrock, ``{The Magnetic Moment of a Massive Neutrino and
  Neutrino Spin Rotation},''
  \href{http://dx.doi.org/10.1103/PhysRevLett.45.963}{{\em Phys. Rev. Lett.}
  {\bfseries 45} (1980) 963}.

\bibitem{Pal:1981rm}
P.~B. Pal and L.~Wolfenstein, ``{Radiative Decays of Massive Neutrinos},''
  \href{http://dx.doi.org/10.1103/PhysRevD.25.766}{{\em Phys. Rev.} {\bfseries
  D25} (1982) 766}.

\bibitem{Shrock:1982sc}
R.~E. Shrock, ``{Electromagnetic properties and decays of Dirac and Majorana
  neutrinos in a general class of gauge theories},''
  \href{http://dx.doi.org/10.1016/0550-3213(82)90273-5}{{\em Nucl. Phys.}
  {\bfseries B206} (1982) 359}.

\bibitem{Bernabeu:2000hf}
J.~Bernabeu, L.~G. Cabral-Rosetti, J.~Papavassiliou, and J.~Vidal, ``{On the
  charge radius of the neutrino},''
  \href{http://dx.doi.org/10.1103/PhysRevD.62.113012}{{\em Phys. Rev.}
  {\bfseries D62} (2000) 113012},
  \href{http://arxiv.org/abs/hep-ph/0008114}{{\ttfamily hep-ph/0008114}}.

\bibitem{Bernabeu:2002nw}
J.~Bernabeu, J.~Papavassiliou, and J.~Vidal, ``{On the observability of the
  neutrino charge radius},'' {\em Phys. Rev. Lett.} {\bfseries 89} (2002)
  101802, \href{http://arxiv.org/abs/hep-ph/0206015}{{\ttfamily
  hep-ph/0206015}}.

\bibitem{Bernabeu:2002pd}
J.~Bernabeu, J.~Papavassiliou, and J.~Vidal, ``{The neutrino charge radius is a
  physical observable},'' {\em Nucl. Phys.} {\bfseries B680} (2004) 450,
  \href{http://arxiv.org/abs/hep-ph/0210055}{{\ttfamily hep-ph/0210055}}.

\bibitem{Cadeddu:2018dux}
M.~Cadeddu, C.~Giunti, K.~A. Kouzakov, Y.~F. Li, A.~I. Studenikin, and Y.~Y.
  Zhang, ``{Neutrino Charge Radii from COHERENT Elastic Neutrino-Nucleus
  Scattering},'' {\em Phys.Rev.} {\bfseries D98} (2018) 113010,
  \href{http://arxiv.org/abs/arXiv:1810.05606}{{\ttfamily arXiv:1810.05606
  [hep-ph]}}.

\bibitem{Cadeddu:2019eta}
M.~Cadeddu, F.~Dordei, C.~Giunti, Y.~F. Li, and Y.~Y. Zhang, ``{Neutrino,
  Electroweak and Nuclear Physics from COHERENT Elastic Neutrino-Nucleus
  Scattering with Refined Quenching Factor},'' {\em Phys.Rev.} {\bfseries D101}
  (2020) 033004, \href{http://arxiv.org/abs/arXiv:1908.06045}{{\ttfamily
  arXiv:1908.06045 [hep-ph]}}.

\bibitem{LZ:2022ysc}
{\bfseries LZ} Collaboration, J.~Aalbers {\em et~al.}, ``{Background
  determination for the LUX-ZEPLIN dark matter experiment},''
  \href{http://dx.doi.org/10.1103/PhysRevD.108.012010}{{\em Phys. Rev. D}
  {\bfseries 108} no.~1, (2023) 012010},
  \href{http://arxiv.org/abs/2211.17120}{{\ttfamily arXiv:2211.17120
  [hep-ex]}}.

\bibitem{LZ:2023poo}
{\bfseries LZ} Collaboration, J.~Aalbers {\em et~al.}, ``{A search for new
  physics in low-energy electron recoils from the first LZ exposure},''
  \href{http://arxiv.org/abs/2307.15753}{{\ttfamily arXiv:2307.15753
  [hep-ex]}}.

\bibitem{A:2022acy}
S.~K. A., A.~Majumdar, D.~K. Papoulias, H.~Prajapati, and R.~Srivastava,
  ``{Implications of first LZ and XENONnT results: A comparative study of
  neutrino properties and light mediators},''
  \href{http://dx.doi.org/10.1016/j.physletb.2023.137742}{{\em Phys. Lett. B}
  {\bfseries 839} (2023) 137742},
  \href{http://arxiv.org/abs/2208.06415}{{\ttfamily arXiv:2208.06415
  [hep-ph]}}.

\bibitem{XENON:2020rca}
{\bfseries XENON} Collaboration, E.~Aprile {\em et~al.}, ``{Excess electronic
  recoil events in XENON1T},''
  \href{http://dx.doi.org/10.1103/PhysRevD.102.072004}{{\em Phys. Rev. D}
  {\bfseries 102} no.~7, (2020) 072004},
  \href{http://arxiv.org/abs/2006.09721}{{\ttfamily arXiv:2006.09721
  [hep-ex]}}.

\bibitem{PandaX-II:2020udv}
{\bfseries PandaX-II} Collaboration, X.~Zhou {\em et~al.}, ``{A Search for
  Solar Axions and Anomalous Neutrino Magnetic Moment with the Complete
  PandaX-II Data},''
  \href{http://dx.doi.org/10.1088/0256-307X/38/10/109902}{{\em Chin. Phys.
  Lett.} {\bfseries 38} no.~1, (2021) 011301},
  \href{http://arxiv.org/abs/2008.06485}{{\ttfamily arXiv:2008.06485
  [hep-ex]}}. [Erratum: Chin.Phys.Lett. 38, 109902 (2021)].

\bibitem{Khan:2022bel}
A.~N. Khan, ``{Light new physics and neutrino electromagnetic interactions in
  XENONnT},'' \href{http://dx.doi.org/10.1016/j.physletb.2022.137650}{{\em
  Phys. Lett. B} {\bfseries 837} (2023) 137650},
  \href{http://arxiv.org/abs/2208.02144}{{\ttfamily arXiv:2208.02144
  [hep-ph]}}.

\bibitem{AtzoriCorona:2022qrf}
M.~Atzori~Corona, M.~Cadeddu, N.~Cargioli, F.~Dordei, C.~Giunti, Y.~F. Li,
  C.~A. Ternes, and Y.~Y. Zhang, ``{Impact of the Dresden-II and COHERENT
  neutrino scattering data on neutrino electromagnetic properties and
  electroweak physics},'' \href{http://dx.doi.org/10.1007/JHEP09(2022)164}{{\em
  JHEP} {\bfseries 09} (2022) 164},
  \href{http://arxiv.org/abs/2205.09484}{{\ttfamily arXiv:2205.09484
  [hep-ph]}}.

\bibitem{Coloma:2022avw}
P.~Coloma, I.~Esteban, M.~C. Gonzalez-Garcia, L.~Larizgoitia, F.~Monrabal, and
  S.~Palomares-Ruiz, ``{Bounds on new physics with data of the Dresden-II
  reactor experiment and COHERENT},''
  \href{http://dx.doi.org/10.1007/JHEP05(2022)037}{{\em JHEP} {\bfseries 05}
  (2022) 037}, \href{http://arxiv.org/abs/2202.10829}{{\ttfamily
  arXiv:2202.10829 [hep-ph]}}.

\bibitem{Liao:2022hno}
J.~Liao, H.~Liu, and D.~Marfatia, ``{Implications of the first evidence for
  coherent elastic scattering of reactor neutrinos},''
  \href{http://dx.doi.org/10.1103/PhysRevD.106.L031702}{{\em Phys. Rev. D}
  {\bfseries 106} no.~3, (2022) L031702},
  \href{http://arxiv.org/abs/2202.10622}{{\ttfamily arXiv:2202.10622
  [hep-ph]}}.

\bibitem{Khan:2022akj}
A.~N. Khan, ``{Neutrino millicharge and other electromagnetic interactions with
  COHERENT-2021 data},''
  \href{http://dx.doi.org/10.1016/j.nuclphysb.2022.116064}{{\em Nucl. Phys. B}
  {\bfseries 986} (2023) 116064},
  \href{http://arxiv.org/abs/2201.10578}{{\ttfamily arXiv:2201.10578
  [hep-ph]}}.

\bibitem{DeRomeri:2022twg}
V.~De~Romeri, O.~G. Miranda, D.~K. Papoulias, G.~Sanchez~Garcia, M.~T\'ortola,
  and J.~W.~F. Valle, ``{Physics implications of a combined analysis of
  COHERENT CsI and LAr data},''
  \href{http://dx.doi.org/10.1007/JHEP04(2023)035}{{\em JHEP} {\bfseries 04}
  (2023) 035}, \href{http://arxiv.org/abs/2211.11905}{{\ttfamily
  arXiv:2211.11905 [hep-ph]}}.

\bibitem{Coloma:2022umy}
P.~Coloma, P.~Coloma, M.~C. Gonzalez-Garcia, M.~C. Gonzalez-Garcia, M.~Maltoni,
  M.~Maltoni, J.~a.~P. Pinheiro, J.~a.~P. Pinheiro, S.~Urrea, and S.~Urrea,
  ``{Constraining new physics with Borexino Phase-II spectral data},''
  \href{http://dx.doi.org/10.1007/JHEP07(2022)138}{{\em JHEP} {\bfseries 07}
  (2022) 138}, \href{http://arxiv.org/abs/2204.03011}{{\ttfamily
  arXiv:2204.03011 [hep-ph]}}. [Erratum: JHEP 11, 138 (2022)].

\bibitem{MUNU:2005xnz}
{\bfseries MUNU} Collaboration, Z.~Daraktchieva {\em et~al.}, ``{Final results
  on the neutrino magnetic moment from the MUNU experiment},''
  \href{http://dx.doi.org/10.1016/j.physletb.2005.04.030}{{\em Phys. Lett. B}
  {\bfseries 615} (2005) 153--159},
  \href{http://arxiv.org/abs/hep-ex/0502037}{{\ttfamily arXiv:hep-ex/0502037}}.

\bibitem{TEXONO:2006xds}
{\bfseries TEXONO} Collaboration, H.~T. Wong {\em et~al.}, ``{A Search of
  Neutrino Magnetic Moments with a High-Purity Germanium Detector at the
  Kuo-Sheng Nuclear Power Station},''
  \href{http://dx.doi.org/10.1103/PhysRevD.75.012001}{{\em Phys. Rev. D}
  {\bfseries 75} (2007) 012001},
  \href{http://arxiv.org/abs/hep-ex/0605006}{{\ttfamily arXiv:hep-ex/0605006}}.

\bibitem{Ahrens:1990fp}
L.~A. Ahrens {\em et~al.}, ``{Determination of electroweak parameters from the
  elastic scattering of muon-neutrinos and anti-neutrinos on electrons},''
  \href{http://dx.doi.org/10.1103/PhysRevD.41.3297}{{\em Phys. Rev. D}
  {\bfseries 41} (1990) 3297--3316}.

\bibitem{Allen:1992qe}
R.~C. Allen {\em et~al.}, ``{Study of electron-neutrino electron elastic
  scattering at LAMPF},'' \href{http://dx.doi.org/10.1103/PhysRevD.47.11}{{\em
  Phys. Rev. D} {\bfseries 47} (1993) 11--28}.

\bibitem{LSND:2001akn}
{\bfseries LSND} Collaboration, L.~B. Auerbach {\em et~al.}, ``{Measurement of
  electron - neutrino - electron elastic scattering},''
  \href{http://dx.doi.org/10.1103/PhysRevD.63.112001}{{\em Phys. Rev. D}
  {\bfseries 63} (2001) 112001},
  \href{http://arxiv.org/abs/hep-ex/0101039}{{\ttfamily arXiv:hep-ex/0101039}}.

\bibitem{DONUT:2001zvi}
{\bfseries DONUT} Collaboration, R.~Schwienhorst {\em et~al.}, ``{A New upper
  limit for the tau - neutrino magnetic moment},''
  \href{http://dx.doi.org/10.1016/S0370-2693(01)00746-8}{{\em Phys. Lett. B}
  {\bfseries 513} (2001) 23--29},
  \href{http://arxiv.org/abs/hep-ex/0102026}{{\ttfamily arXiv:hep-ex/0102026}}.

\bibitem{Canas:2015yoa}
B.~C. Canas, O.~G. Miranda, A.~Parada, M.~Tortola, and J.~W.~F. Valle,
  ``{Updating neutrino magnetic moment constraints},''
  \href{http://dx.doi.org/10.1016/j.physletb.2015.12.011}{{\em Phys. Lett. B}
  {\bfseries 753} (2016) 191--198},
  \href{http://arxiv.org/abs/1510.01684}{{\ttfamily arXiv:1510.01684
  [hep-ph]}}. [Addendum: Phys.Lett.B 757, 568--568 (2016)].

\bibitem{AristizabalSierra:2020zod}
D.~Aristizabal~Sierra, R.~Branada, O.~G. Miranda, and G.~Sanchez~Garcia,
  ``{Sensitivity of direct detection experiments to neutrino magnetic dipole
  moments},'' \href{http://dx.doi.org/10.1007/JHEP12(2020)178}{{\em JHEP}
  {\bfseries 12} (2020) 178}, \href{http://arxiv.org/abs/2008.05080}{{\ttfamily
  arXiv:2008.05080 [hep-ph]}}.

\bibitem{Ayala:1998qz}
A.~Ayala, J.~C. D'Olivo, and M.~Torres, ``{Bound on the neutrino magnetic
  moment from chirality flip in supernovae},''
  \href{http://dx.doi.org/10.1103/PhysRevD.59.111901}{{\em Phys. Rev. D}
  {\bfseries 59} (1999) 111901},
  \href{http://arxiv.org/abs/hep-ph/9804230}{{\ttfamily arXiv:hep-ph/9804230}}.

\bibitem{Viaux:2013lha}
N.~Viaux, M.~Catelan, P.~B. Stetson, G.~Raffelt, J.~Redondo, A.~A.~R. Valcarce,
  and A.~Weiss, ``{Neutrino and axion bounds from the globular cluster M5 (NGC
  5904)},'' \href{http://dx.doi.org/10.1103/PhysRevLett.111.231301}{{\em Phys.
  Rev. Lett.} {\bfseries 111} (2013) 231301},
  \href{http://arxiv.org/abs/1311.1669}{{\ttfamily arXiv:1311.1669
  [astro-ph.SR]}}.

\bibitem{Corsico:2014mpa}
A.~H. C\'orsico, L.~G. Althaus, M.~M. Miller~Bertolami, S.~O. Kepler, and
  E.~Garc\'\i{}a-Berro, ``{Constraining the neutrino magnetic dipole moment
  from white dwarf pulsations},''
  \href{http://dx.doi.org/10.1088/1475-7516/2014/08/054}{{\em JCAP} {\bfseries
  08} (2014) 054}, \href{http://arxiv.org/abs/1406.6034}{{\ttfamily
  arXiv:1406.6034 [astro-ph.SR]}}.

\bibitem{Capozzi:2020cbu}
F.~Capozzi and G.~Raffelt, ``{Axion and neutrino bounds improved with new
  calibrations of the tip of the red-giant branch using geometric distance
  determinations},'' \href{http://dx.doi.org/10.1103/PhysRevD.102.083007}{{\em
  Phys. Rev. D} {\bfseries 102} no.~8, (2020) 083007},
  \href{http://arxiv.org/abs/2007.03694}{{\ttfamily arXiv:2007.03694
  [astro-ph.SR]}}.

\bibitem{Mori:2020qqd}
K.~Mori, M.~Kusakabe, A.~B. Balantekin, T.~Kajino, and M.~A. Famiano,
  ``{Enhancement of Lithium in Red Clump Stars by the Additional Energy Loss
  Induced by New Physics},''
  \href{http://dx.doi.org/10.1093/mnras/stab595}{{\em Mon. Not. Roy. Astron.
  Soc.} {\bfseries 503} no.~2, (2021) 2746--2753},
  \href{http://arxiv.org/abs/2009.00293}{{\ttfamily arXiv:2009.00293
  [astro-ph.SR]}}.

\bibitem{TEXONO:2002pra}
{\bfseries TEXONO} Collaboration, H.~B. Li {\em et~al.}, ``{Limit on the
  electron neutrino magnetic moment from the Kuo-Sheng reactor neutrino
  experiment},'' \href{http://dx.doi.org/10.1103/PhysRevLett.90.131802}{{\em
  Phys. Rev. Lett.} {\bfseries 90} (2003) 131802},
  \href{http://arxiv.org/abs/hep-ex/0212003}{{\ttfamily arXiv:hep-ex/0212003}}.

\bibitem{Gninenko:2006fi}
S.~N. Gninenko, N.~V. Krasnikov, and A.~Rubbia, ``{Search for millicharged
  particles in reactor neutrino experiments: A Probe of the PVLAS anomaly},''
  \href{http://dx.doi.org/10.1103/PhysRevD.75.075014}{{\em Phys. Rev. D}
  {\bfseries 75} (2007) 075014},
  \href{http://arxiv.org/abs/hep-ph/0612203}{{\ttfamily arXiv:hep-ph/0612203}}.

\bibitem{Chen:2014dsa}
J.-W. Chen, H.-C. Chi, H.-B. Li, C.~P. Liu, L.~Singh, H.~T. Wong, C.-L. Wu, and
  C.-P. Wu, ``{Constraints on millicharged neutrinos via analysis of data from
  atomic ionizations with germanium detectors at sub-keV sensitivities},''
  \href{http://dx.doi.org/10.1103/PhysRevD.90.011301}{{\em Phys. Rev. D}
  {\bfseries 90} no.~1, (2014) 011301},
  \href{http://arxiv.org/abs/1405.7168}{{\ttfamily arXiv:1405.7168 [hep-ph]}}.

\bibitem{Studenikin:2013my}
A.~Studenikin, ``{New bounds on neutrino electric millicharge from limits on
  neutrino magnetic moment},''
  \href{http://dx.doi.org/10.1209/0295-5075/107/21001}{{\em EPL} {\bfseries
  107} no.~2, (2014) 21001}, \href{http://arxiv.org/abs/1302.1168}{{\ttfamily
  arXiv:1302.1168 [hep-ph]}}. [Erratum: EPL 107, 39901 (2014), Erratum:
  Europhys.Lett. 107, 39901 (2014)].

\bibitem{XMASS:2020zke}
{\bfseries XMASS} Collaboration, K.~Abe {\em et~al.}, ``{Search for exotic
  neutrino-electron interactions using solar neutrinos in XMASS-I},''
  \href{http://dx.doi.org/10.1016/j.physletb.2020.135741}{{\em Phys. Lett. B}
  {\bfseries 809} (2020) 135741},
  \href{http://arxiv.org/abs/2005.11891}{{\ttfamily arXiv:2005.11891
  [hep-ex]}}.

\bibitem{Raffelt:1999gv}
G.~G. Raffelt, ``{Limits on neutrino electromagnetic properties: An update},''
  \href{http://dx.doi.org/10.1016/S0370-1573(99)00074-5}{{\em Phys. Rept.}
  {\bfseries 320} (1999) 319--327}.

\bibitem{TEXONO:2009knm}
{\bfseries TEXONO} Collaboration, M.~Deniz {\em et~al.}, ``{Measurement of
  Nu(e)-bar -Electron Scattering Cross-Section with a CsI(Tl) Scintillating
  Crystal Array at the Kuo-Sheng Nuclear Power Reactor},''
  \href{http://dx.doi.org/10.1103/PhysRevD.81.072001}{{\em Phys. Rev. D}
  {\bfseries 81} (2010) 072001},
  \href{http://arxiv.org/abs/0911.1597}{{\ttfamily arXiv:0911.1597 [hep-ex]}}.

\bibitem{CHARM-II:1994aeb}
{\bfseries CHARM-II} Collaboration, P.~Vilain {\em et~al.}, ``{Experimental
  study of electromagnetic properties of the muon-neutrino in neutrino -
  electron scattering},''
  \href{http://dx.doi.org/10.1016/0370-2693(94)01678-6}{{\em Phys. Lett. B}
  {\bfseries 345} (1995) 115--118}.

\bibitem{Khan:2017djo}
A.~N. Khan, ``{$\sin ^{2}\theta _{W}$ Estimate and Neutrino Electromagnetic
  Properties from Low-Energy Solar Data},''
  \href{http://dx.doi.org/10.1088/1361-6471/ab0057}{{\em J. Phys. G} {\bfseries
  46} no.~3, (2019) 035005}, \href{http://arxiv.org/abs/1709.02930}{{\ttfamily
  arXiv:1709.02930 [hep-ph]}}.

\end{thebibliography}

\providecommand{\href}[2]{#2}\begingroup\raggedright\endgroup

\end{document}